\newif\ifnotes
\notesfalse

\newif\iftwocol
\twocolfalse

\iftwocol
  \documentclass[conference]{IEEEtran}
\else
  \documentclass[letterpaper,11pt,english]{article}
  \usepackage[verbose,tmargin=3cm,bmargin=3cm,lmargin=3cm,rmargin=3cm]{geometry}
\fi

\newif\ifusenix
\usenixfalse

\newcommand{\full}[1]{}

\newcommand{\LONGBIB}[1]{}

\usepackage[T1]{fontenc}
\usepackage[latin9]{inputenc}

\usepackage{epsfig}
\usepackage{endnotes}
\usepackage{balance}
\usepackage{url}
\usepackage{algorithm}
\usepackage[noend]{algpseudocode}

\usepackage{amsmath}
\usepackage{listings}
\usepackage{amsthm}
\usepackage{color}
\usepackage{pbox}

\usepackage{graphicx}
\usepackage{tabularx}
\usepackage{comment}
\usepackage{subfig}

\usepackage{enumitem}
\date{\today\enspace \small{(initial publication September 11, 2018)}}

\iftwocol
\else
  \usepackage[numbers]{natbib}
  \usepackage[font=small]{caption}
\fi

\usepackage{etoolbox}
\patchcmd{\thebibliography}{\chapter*}{\section*}{}{}

\makeatletter

\def\BState{\State\hskip-\ALG@thistlm}

\numberwithin{equation}{section}
\numberwithin{figure}{section}

\usepackage{url}

\definecolor{darkgreen}{rgb}{0,0.6,0}
\newcommand{\eran}[1]{\dtcolornote[Eran]{darkgreen}{#1}}
\newcommand{\roei}[1]{\dtcolornote[Roei]{blue}{#1}}
\newcommand{\daniel}[1]{\dtcolornote[Daniel]{orange}{#1}}

\expandafter\def\expandafter\UrlBreaks\expandafter{\UrlBreaks%COMMENT
  \do\a\do\b\do\c\do\d\do\e\do\f\do\g\do\h\do\i\do\j%COMMENT
  \do\k\do\l\do\m\do\n\do\o\do\p\do\q\do\r\do\s\do\t%COMMENT
  \do\u\do\v\do\w\do\x\do\y\do\z\do\A\do\B\do\C\do\D%COMMENT
  \do\E\do\F\do\G\do\H\do\I\do\J\do\K\do\L\do\M\do\N%COMMENT
  \do\O\do\P\do\Q\do\R\do\S\do\T\do\U\do\V\do\W\do\X%COMMENT
  \do\Y\do\Z}

\newif\iffull
 \fulltrue
\fullfalse

\ifnotes
  \usepackage[notes=true, done=false, later=false, draft]{dtrt}
\else
  \usepackage[notes=xxx]{dtrt}
\fi

\usepackage{sanebib}

\usepackage{etoolbox}
\makeatletter
\patchcmd{\ttlh@hang}{\parindent\z@}{\parindent\z@\leavevmode}{}{}
\patchcmd{\ttlh@hang}{\noindent}{}{}{}
\usepackage{xspace}
\newcommand{\preunit}{\unskip\,}

\newcommand{\Hz}{\preunit{\text{Hz}}\xspace}
\newcommand{\kHz}{\preunit{\text{kHz}}\xspace}

\newcommand{\MHz}{\preunit{\text{MHz}}\xspace}

\newcommand{\V}{\preunit{\text{V}}\xspace}
 
\newcommand{\second}{\preunit{\text{s}}\xspace} 
\newcommand{\kSps}{\preunit{\text{kHz}}\xspace}  %COMMENT
\newcommand{\meter}{\preunit{\text{m}}\xspace}

\begin{document}

\title{
	\iftwocol{}
	Synesthesia: Detecting Screen Content \iftwocol\else \\ \fi via Remote Acoustic Side Channels\textsuperscript{\small{*}}
\vspace{-1.3em}
\else{}
  Synesthesia: Detecting Screen Content \iftwocol\else \\ \fi via Remote Acoustic Side Channels%COMMENT
  \footnote{
    Roei Schuster and Eran Tromer are members of the Check Point Institute for Information Security.\newline
    This paper is forthcoming in IEEE Security and Privacy 2019, DOI 10.1109/SP.2019.00074.\newline
    Authors are ordered alphabetically.
  }
  \fi{}
}
\hypersetup{pdftitle={Synesthesia: Detecting Screen Content via Remote Acoustic Side Channels}}

%COMMENT
%COMMENT
\begin{comment}
%COMMENT
\makeatletter
\def\and{%COMMENT
  \end{tabular}%COMMENT
  \hskip 0.9em \@plus.17fil\relax
  \begin{tabular}[t]{c}}
\makeatother
\end{comment}

%COMMENT
%COMMENT
%COMMENT
%COMMENT
%COMMENT
\author{
{\rm Daniel Genkin}\\
{\normalsize{}University of Michigan}\\
{\normalfont\texttt{\normalsize{genkin@umich.edu}}}
\and
{\rm Mihir Pattani}\\
{\normalsize{}University of Pennsylvania}\\
{\normalfont\texttt{\normalsize{}mihirsa@seas.upenn.edu}}
\and
{\rm Roei Schuster}\\
{\normalsize{}Tel Aviv University,
\normalsize{}Cornell Tech}\\
{\normalfont\texttt{\normalsize{}rs864@cornell.edu}}
\and
{\rm Eran Tromer}\\
{\normalsize{}Tel Aviv University,
\normalsize{}Columbia University}\\
{\normalfont\texttt{\normalsize{}tromer@cs.tau.ac.il}}
}
%COMMENT

%COMMENT
%COMMENT
%COMMENT
%COMMENT
%COMMENT
%COMMENT
%COMMENT
%COMMENT
%COMMENT
%COMMENT
%COMMENT
%COMMENT
%COMMENT
%COMMENT
%COMMENT

\maketitle

\iftwocol{}
\renewcommand*{\thefootnote}{\fnsymbol{footnote}}

\setcounter{footnote}{1}

\footnotetext{
    Authors are ordered alphabetically.
  }
\renewcommand*{\thefootnote}{\arabic{footnote}}
\setcounter{footnote}{0}
\else{}
\fi{}

\begin{abstract}
We show that subtle acoustic noises emanating from within computer screens can be used to detect the content displayed on the screens. This sound can be picked up by ordinary microphones built into webcams or screens, and is inadvertently transmitted to other parties, e.g., during a videoconference call or archived recordings.
It can also be recorded by a smartphone or ``smart speaker'' placed on a desk next to the screen, or from as far as 10 meters away using a parabolic microphone. 

Empirically demonstrating various attack scenarios, we show how this channel can be used for real-time detection of on-screen text, or users' input into on-screen virtual keyboards.  We also demonstrate how an attacker can analyze the audio received during video call (e.g., on Google Hangout) to infer whether the other side is browsing the web in lieu of watching the video call, and which web site is displayed on their screen.
\end{abstract}

%COMMENT
%COMMENT

%COMMENT
%COMMENT
%COMMENT

\makeatother

%COMMENT

\section{Introduction}
\label{sec:intro}
Physical side-channel attacks extract information from computing systems by measuring unintended effects of a system on its physical environment. 
%COMMENT
They have been used to violate the security of numerous cryptographic implementations (see~\cite{Power-analysis-smart-cards-book, Anderson08seceng2nd,kjjr11introduction} and the references therein), both on small embedded devices and, more recently, on complex devices such as laptops, PCs, and smartphones~\cite{CZP15,GPST16,BFMT16,GPPTY16ecdsa, clark2013wattsupdoc, clark2013current, GPT14eprintWconf}.
 \done\roei{SANITIZED}
Physical emanations were used to recover information from peripheral input/output devices such as
screens~\cite{Eck85electromagneticradiation,Kuhn04, enev2011televisions}, printers~\cite{BDGPS10} and keyboards~\cite{AA04,BWY06,vuagnoux2009compromising, halevi2012closer, balzarotti2008clearshot, halevi2015keyboard, zhuang2009keyboard,CCLT17skypetype}.
\ignore[inexpensive] {
  Some of these attacks can be conducted using inexpensive equipment, including some attacks on cryptography~\cite{GPPT15, GPT14eprintWconf, GST14acoustic-norefs} and the acoustic attacks on keyboards and printers.
}

Attackers seeking to exploit such channels face a challenge: attaining physical proximity to the target computer, in order to acquire physical measurements. In many settings, physical access is controlled, and attempts to attain proximity will be blocked or detected. Alternatively, the attacker can seek to control suitable sensors that are already located in close proximity to the target; this may be tractable when there are ubiquitously deployed commodity devices, with suitable sensors, that can be adversarially controlled; for example, one of the low-bandwidth acoustic attacks~\cite{GST14acoustic-norefs} can be conducted via a smartphone, using a malicious app that records audio using the built-in microphone when the smartphone is placed (for an hour) near the target.

We raise a third possibility: are there physical side-channel attacks for which the requisite physical measurements are readily available, and are shared by victims with untrusted parties as a matter of course, and yet the victims have no reason to suspect that they inadvertently leak private information?

We observe a new physical side channel that facilitates such an attack: \emph{content-dependent acoustic leakage from LCD screens}. This leakage can be picked up by adjacent microphones, such as those embedded in webcams and some computer screens. Users commonly share audio recorded by these microphones, e.g., during Voice over IP and videoconference calls. Moreover, the pertinent sounds are so faint and high-pitched that they are well-nigh inaudible to the human ear, and thus (unlike with mechanical peripherals) users have no reason to suspect that these emanations exist and that information about their screen content is being conveyed to anyone who receives the audio stream, or even a retroactive recording. In fact, users often make an effort to place their webcam (and thus, microphone) in close proximity to the screen, in order to maintain eye contact during videoconference, thereby offering high quality measurements to would-be attackers.

Exploiting this channel raises many questions: What form of content-dependent acoustic signals are emitted by screens? Can these emanations be used to detect screen content? What measurement equipment and positioning suffices to measure them? Can they be acquired remotely via Voice over IP applications, despite the signal conditioning and lossy codecs employed by such applications?

\subsection{Our results}

We observe the existence of the aforementioned synesthetic side channel: ``hearing'' on-screen images. 
We characterize this content-dependent acoustic leakage on numerous LCD screens of various models and manufacturers. The leakage has existed in screens manufactured and sold for at least the past 16 years, old and new models alike, in both PC and laptop screens, with both CCFL and LED backlighting. See Appendix~\ref{sec:screenlist} for a list of screens we experimented with, all of them were found to exhibit this acoustic leakage.
\done\eran{SANITIZED}

\noindent
We show that this leakage can be captured by:
\iftwocol{}
\begin{itemize}[nolistsep, leftmargin=*]
\else{}
\begin{itemize}
\fi{}
  \item Webcam microphones (see Figure~\ref{fig:webCams}).
  \item Mobile phones in proximity to the screen (Figures~\ref{phones}).
  \item ``Smart speaker'' virtual assistant devices (Figure~\ref{fig:alexa}).
  \item The built-in microphones of some screens.
  \item A parabolic microphone from a 10-meters line-of-sight to the back of the screen (see Figure~\ref{fig:parabola}).
\end{itemize}

\noindent
Moreover, the leakage can be observed and analyzed:
\iftwocol{}
\begin{itemize}[nolistsep, leftmargin=*]
\else{}
\begin{itemize}
\fi{}
  \item In archived audio recordings.
  \item From the remote side of a Google Hangouts video-conference call.
  \item In the cloud-stored audio saved by virtual assistants.
\end{itemize}

%COMMENT
%COMMENT
%COMMENT

\noindent
We demonstrate exploitation of the above attack vectors for several attack goals: 
\iftwocol{}
\begin{enumerate}[nolistsep, leftmargin=*]
\else{}
\begin{enumerate}
\fi{}
  \item Extracting text displayed on the screen (in a large font).
  \item Distinguishing between websites displayed on the screen, or between websites and a videoconference screen.
  \item Extracting text entered via Ubuntu's on-screen keyboard. (This shows that on-screen keyboards are not an acoustic-leakage-resistant alternative to mechanical keyboards.)
\end{enumerate}

\noindent Our attacks use tailored signal processing combined with deep neural networks. We demonstrate that leakage characteristics, and trained neural-networks, often generalize across screens (even of different models) allowing for training to be done on one screen while attacking another.  Finally, the attacks are robust to office-environment noise-level from nearby equipment such as computers, other monitors, and human speech.

\subsection{Related work}

\parhead{Physical side channels}
Numerous prior works demonstrated physical side channels. Categorized by channels, these include 
electromagnetic radiation~\cite{quisquaterElectromagnetic,Gandolfi01electromagneticanalysis,   Eck85electromagneticradiation,Kuhn04, enev2011televisions};
power consumption~\cite{Kocher99differentialpower,kjjr11introduction,Power-analysis-smart-cards-book};
ground-potential fluctuations~\cite{GPT14eprintWconf};
timing (often observable by any of the above)~\cite{Kocher96,BB05,BT11};
and acoustic emanations from keyboards~\cite{AA04,BWY06,vuagnoux2009compromising, halevi2012closer, balzarotti2008clearshot, halevi2015keyboard, zhuang2009keyboard,CCLT17skypetype}, printers~\cite{BDGPS10} and CPU power supplies~\cite{GST14acoustic-norefs}. Acoustic emanations are also mentioned in NACSIM~5000 ``TEMPEST Fundamentals''~\cite{NACSIM5000fundam} and related US government publications, but only in the context of electromechanical devices (as described in \cite{NACSIM5000fundam}: ``[...] mechanical operations occur and sound is produced. Keyboards, printers, relays --- these produce sound, and consequently can be sources of compromise''; see~\cite{GST14acoustic-norefs} for further discussion).

While some physical side channels have been thoroughly explored, the only previous work extracting information from involuntary acoustic leakage of electronic components is Genkin et al.'s acoustic cryptanalysis~\cite{GST14acoustic-norefs}, which exploits laptop coil whine, and does not consider leakage from displays.

\parhead{Screen emanations}
Extracting screen content via electromagnetic emanations (``Van Eck phreaking'' and screen ``TEMPEST'') is well known and studied, originally for CRT screens~\cite{Eck85electromagneticradiation}, and later also for modern flat-panel screens and digital interfaces~\cite{Kuhn04,KuhnPHD}. %COMMENT
Such electromagnetic attacks require antennas and radio receivers in physical proximity to the screen, and tuned to suitable radio frequencies. Acoustic attacks relying on microphones, which are ubiquitous and open new attack scenarios, have not been previously addressed.

\subsection{Outline}
This paper is organized as follows: Section~\ref{characterizing} characterizes the content-dependent acoustic leakage from various screens, and suggests a signal processing scheme for producing clean leakage traces. Section~\ref{attackvectors} introduces the attack vectors evaluated in this paper. We then\full{SANITIZED} discuss several attack scenarios, categorized by the attacker's goal: an on-screen keyboard snooping attack (Section~\ref{keyboardattack}), a text extraction attack (Section~\ref{textattack}), and a website-distinguishing attack from various vantage points, including a remote attack over a Hangouts VoIP call (Section~\ref{websitedistinguish}). Section~\ref{sec:crossscreenall} explores generalization of leakage characteristics, as modeled by the attacker, across screens and screen models. Sections~\ref{sec:limitations} discusses limitations, Section~\ref{sec:mitigations} discusses mitigations, and Section~\ref{conclusions} concludes.

\section{Characterizing the\full{SANITIZED} signal}
\label{characterizing}
In this section we explore the acoustic leakage signal emitted by various LCD computer screens (with both CCFL and LED backlighting) as well as attempt to characterize the connections between the leakage signal and the displayed image. We begin by providing some background about the process of rendering an image on modern LCD screens as well as by describing the experimental setup used in this section.

\subsection{Background and experimental setup}
\label{background}
\parhead{Image rendering mechanism} Computer screens display a rectangular $l\times n$ matrix of pixels. Each pixel is typically further divided into red, green, and blue sub-pixels where the intensity of each sub-pixel is an integer between 0 and 255, thereby allowing the color of pixel to be uniquely represented using 24-bit integers.   The screen's \textit{refresh rate}, $r$, determines how many times per second an image to be displayed is sent to the screen by the computer's graphics card. The screen then renders the received image by iterating over the pixel rows from top to bottom, with the value of the pixels in each row rendered from left to right. Notice that the
 screen always re-renders the image it displays (using the above-described method) $r$ times per second, even if no changes occurred in the image to be displayed. Typically, screens are refreshed approximately 30, 60, or 120 times per second, with a refresh rate of approximately 60\Hz being the most common.
 
\parhead{Experimental setup} For experiments performed in this section we captured the acoustic noise emanating from various monitors using a Br{\"u}el \& Kjaer 4190 microphone capsule (effective up to approx. 40\kHz, well beyond its nominal 20\kHz range), attached to a Br{\"u}el \& Kjaer 2669 preamplifier. For power supply and further amplification, these were connected to a Br{\"u}el \& Kjaer 2610 or 5935 amplifier. The resulting amplified signal was filtered as necessary using a high-pass filter
(with a cut off frequency between 10 and 21\kHz) and digitized using a USB sound card, at a sampling rate of 192\kSps (either a Creative E-MU 0404 or Focusrite Scarlett 2i2). We visualize the spectrograms of these signals using the Baudline software and (from Section~\ref{sec:timedomainanalysis} onward) process them via custom scripts.

\parhead{Vsync probe} For the purpose of signal exploration, we shall sometimes utilize a trigger signal marking the start of each screen refresh period; this allows us to focus on the acoustic leakage from each screen refresh, separately from the task of synchronizing to the refresh rate. We implement this trigger by constructing a VGA tap cable that exposes the VGA's vsync line which (as specified in the VGA standard) carries a short 0.5~Volt pulse at the start of each screen refresh cycle. Note that the vsync probe is a didactic aid; it is \emph{not} necessary for the actual attacks.

\parhead{Screen selection} In this paper we analyze acoustic leakage present from 31 screens of 12 different models from 6 different manufacturers, with various resolutions and backlight types (LED or CCFL), as tabulated in Appendix~\ref{sec:screenlist}. The screens used are those already present in our university lab and offices at the time of writing (which, heuristically, offers a sample of popular models), augmented by ad hoc purchases of current models sold on Amazon.com. For in-depth investigations throughout the paper we chose, from the above, the screens that presented the clearest signal. For cross-screen experiments we purchased additional screens, chosen primarily by ready availability on eBay of used instances with diverse usage histories.
%COMMENT

\subsection{Exploring the leakage signal}
\label{exploringleakage}
\parhead{Distinguishing various images}
We begin our analysis of acoustic leakage emitted from LCD computer monitors by attempting to distinguish simple repetitive images displayed on the target monitor. For this purpose, we created a simple program that displays patterns of alternating horizontal black and white stripes of equal thickness (in pixels), which we shall refer to 
as \emph{Zebras}. The \emph{period} of a Zebra is the distance, in pixels, between two adjacent black stripes. Finally, we recorded the sound emitted by a Soyo DYLM2086 screen while displaying different such Zebras. See Figure~\ref{fig:monitor-setup-first} for a picture of our recording setup.  

As can be seen from Figure~\ref{fig:monitor-single}, the alternating the displayed Zebras causes clear changes in the monitor's acoustic signature. See Figure~\ref{fig:monitor-many} for additional examples of monitors that displayed particular clear signals. Beyond those presented in figures throughout the paper, we experimented with dozens of other monitors, including both old and newer-generation monitors of various sizes, and observed similar effects. We conclude that the leakage has existed in screens manufactured and sold for at least the past 16 years, old and new models alike, including LED and CCFL-based screens.

\begin{figure}[t]
\includegraphics[width=\columnwidth]{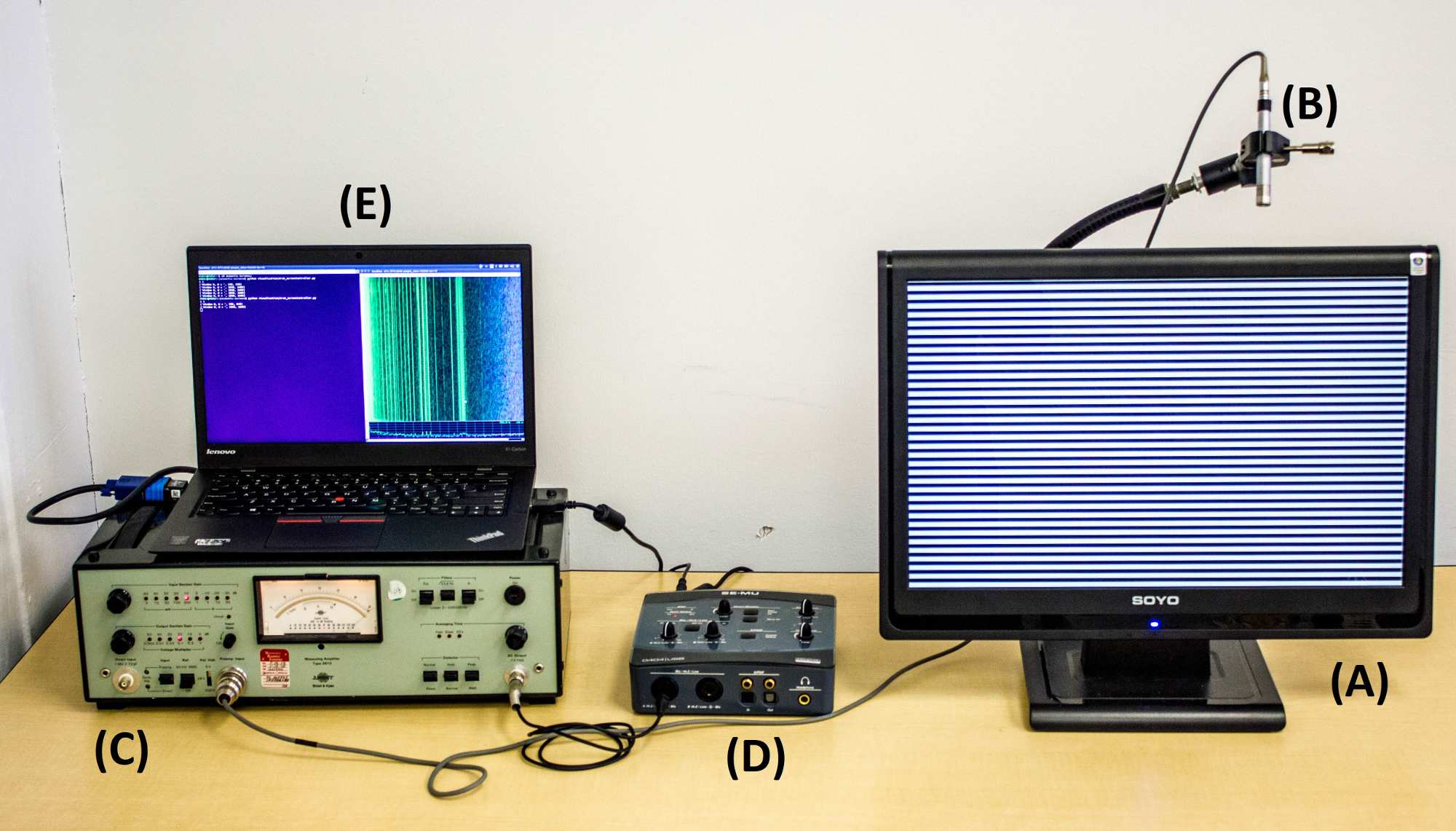}
%COMMENT
\caption{
\label{fig:monitor-setup-first}
Setup used for signal characterization. In this photograph, (A) is a Soyo DYLM2086 target screen, (B) is a Br{\"u}el \& Kjaer 4190 microphone, connected to a Br{\"u}el \& Kjaer 2669 preamplifier, (C) is a Br{\"u}el \& Kjaer 2610 amplifier, (D) is a Creative E-MU 0404 USB sound card with a 192\kSps sample rate (E) is a laptop performing the attack.
}
\iftwocol{}
\vspace{-2em}
\fi{}
\end{figure}

{
\begin{figure}[t]
\makeatletter
\renewcommand{\p@subfigure}{\thefigure--}
\makeatother
\hfil
\subfloat[covered microphone]{
	\iftwocol{}
	\includegraphics[height=0.135\textwidth]{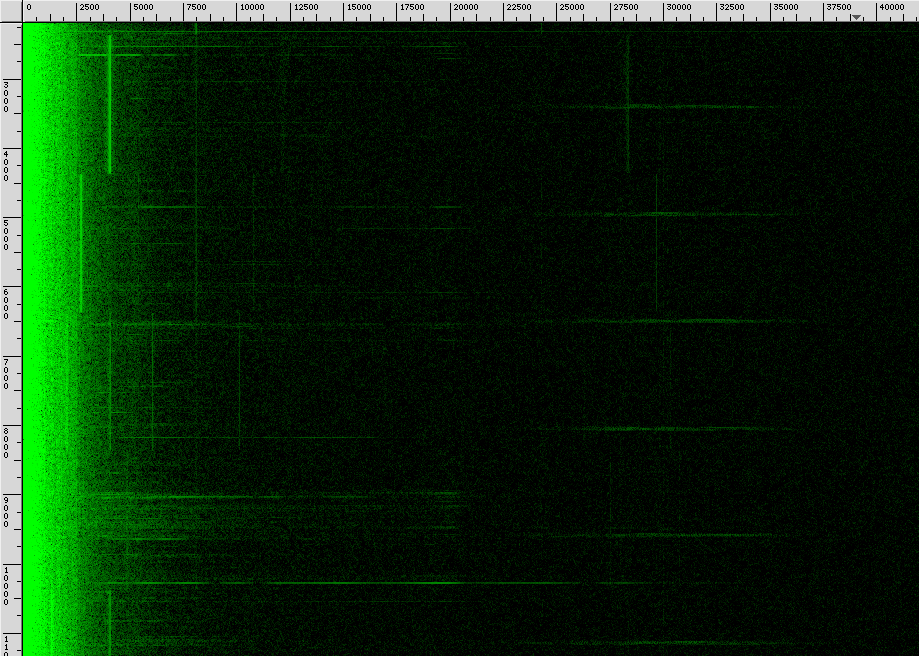}
	\else{}
	\includegraphics[width=0.42\textwidth]{art/daniels_images_final/soyo_s_05_bk_hanky.png}
	\fi{}
	\label{fig:monitor-single-covered}}
\hfil
\subfloat[uncovered microphone]{
	\label{fig:monitor-single}
	\iftwocol{}
	\includegraphics[height=0.135\textwidth]{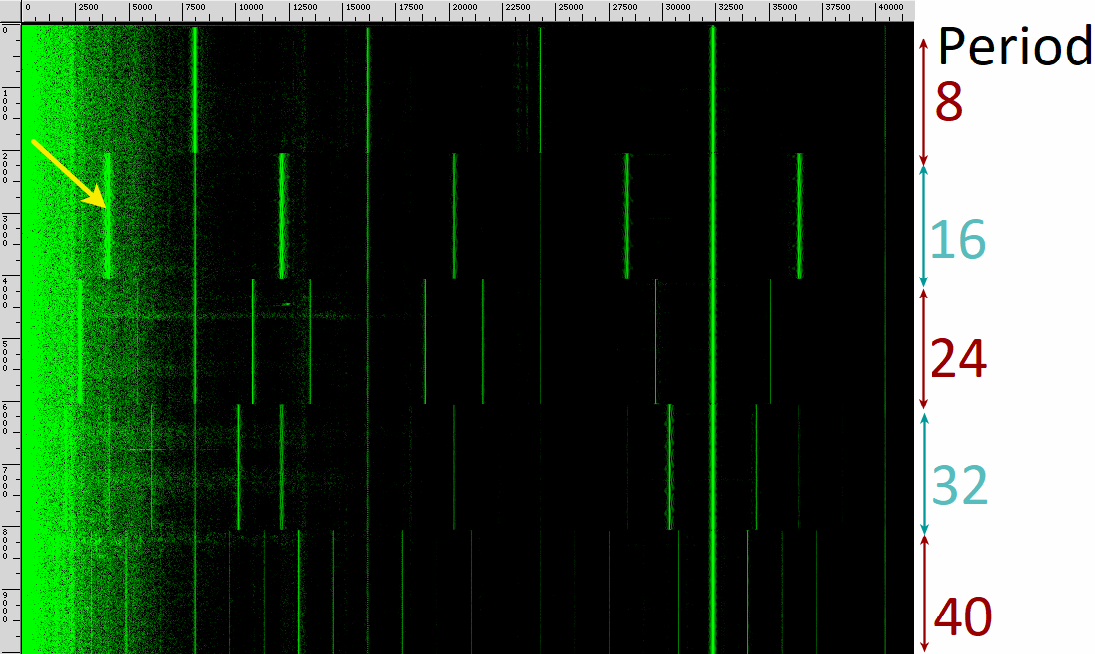}
	\else{}
	\includegraphics[width=0.5\textwidth]{art/daniels_images_final/soyo_s_05_bk_with_arrow.png}
	\fi{}
	}
\hfil
%COMMENT
%COMMENT
%COMMENT
\caption{
	\label{fig:monitor-single-general} (right)
	Spectrogram of acoustic signals emitted by Soyo DYLM2086 screen while displaying alternating Zebra patterns. Notice the difference in the screen's acoustic signature caused by the change in periods of the displayed Zebra pattern. (left) Spectrogram of a recording in an identical setting, but with the microphone covered by a thick piece of cloth. In both spectrograms, the horizontal axis is frequency (0-43 kHz), the vertical
axis is time (10 sec), and intensity is proportional to instantaneous energy in that frequency band. The yellow arrow marks the 4\kHz leakage signal expected to be produced by a Zebra pattern with a period of 16 pixels. 
} 
\end{figure}
}

{
\begin{figure*}[t]
		\makeatletter
		\renewcommand{\p@subfigure}{\thefigure--}
		\makeatother
		\subfloat[Samsung 920NW]{
			\includegraphics[width=0.24\textwidth]{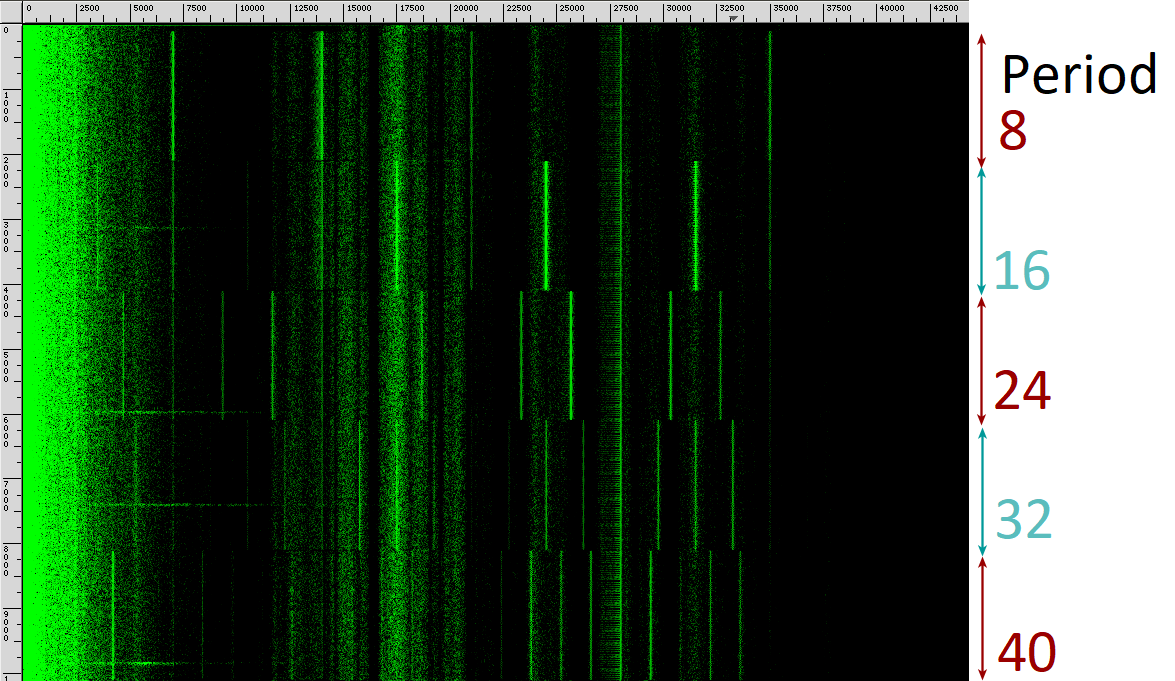}
			%COMMENT
		}
		\subfloat[HP ZR30w]{
		\includegraphics[width=0.245\textwidth]{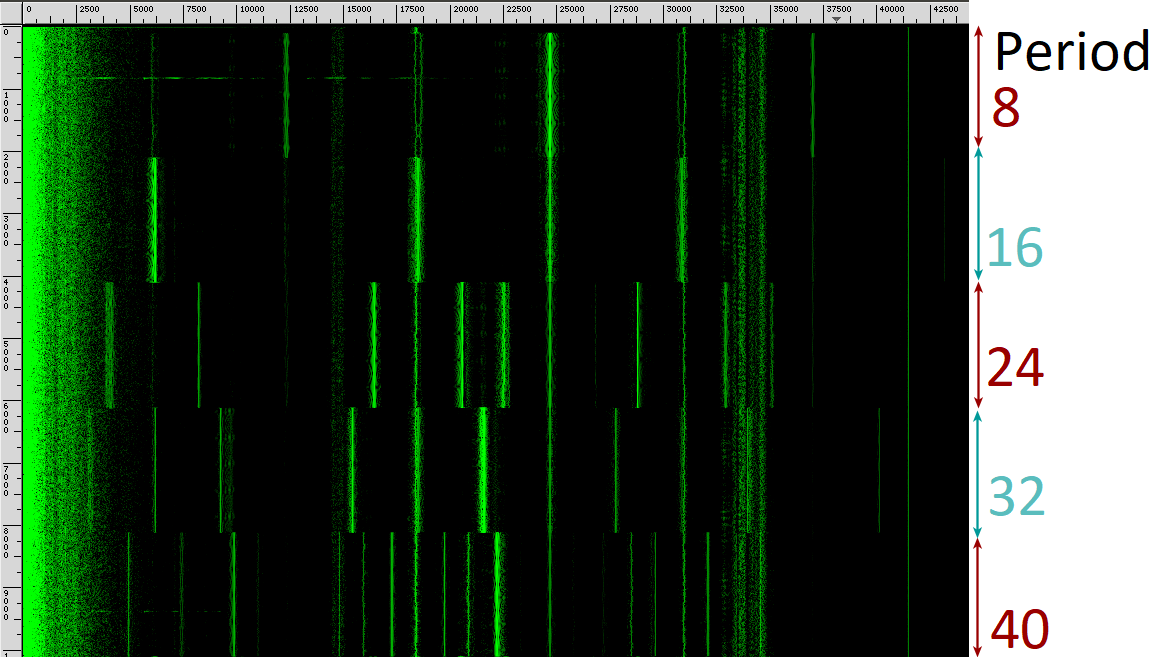}
	%COMMENT
		}
\subfloat[Dell U3011t]{
	\includegraphics[width=0.245\textwidth]{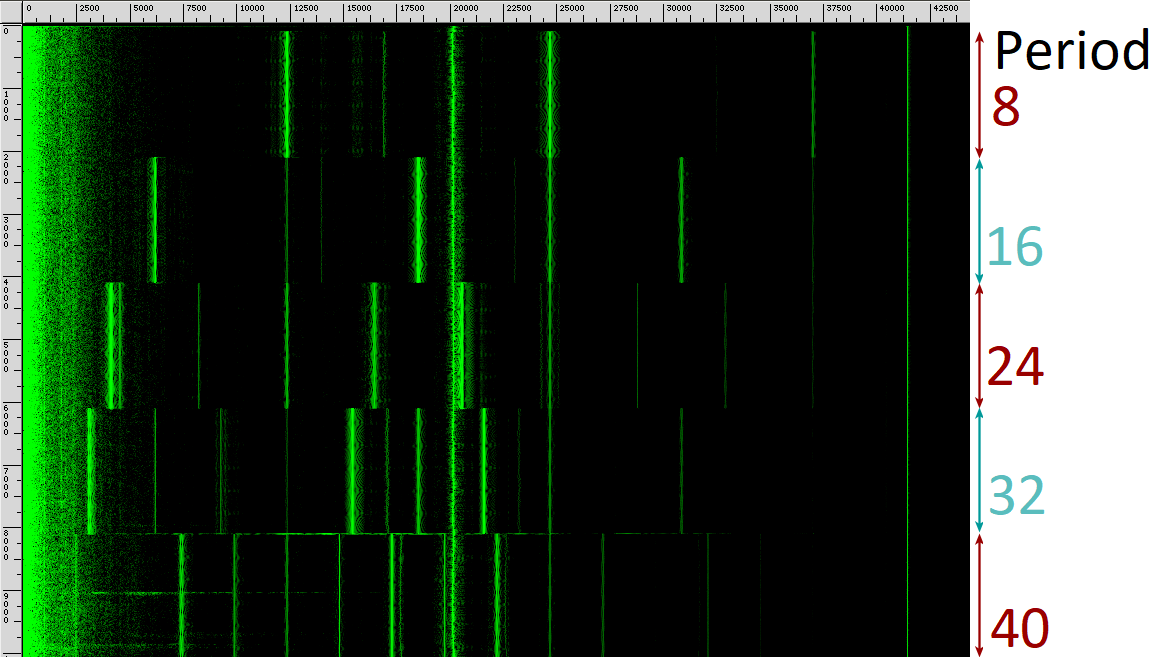}
	%COMMENT
}
				\subfloat[Philips 170S4]{
		\includegraphics[width=0.245\textwidth]{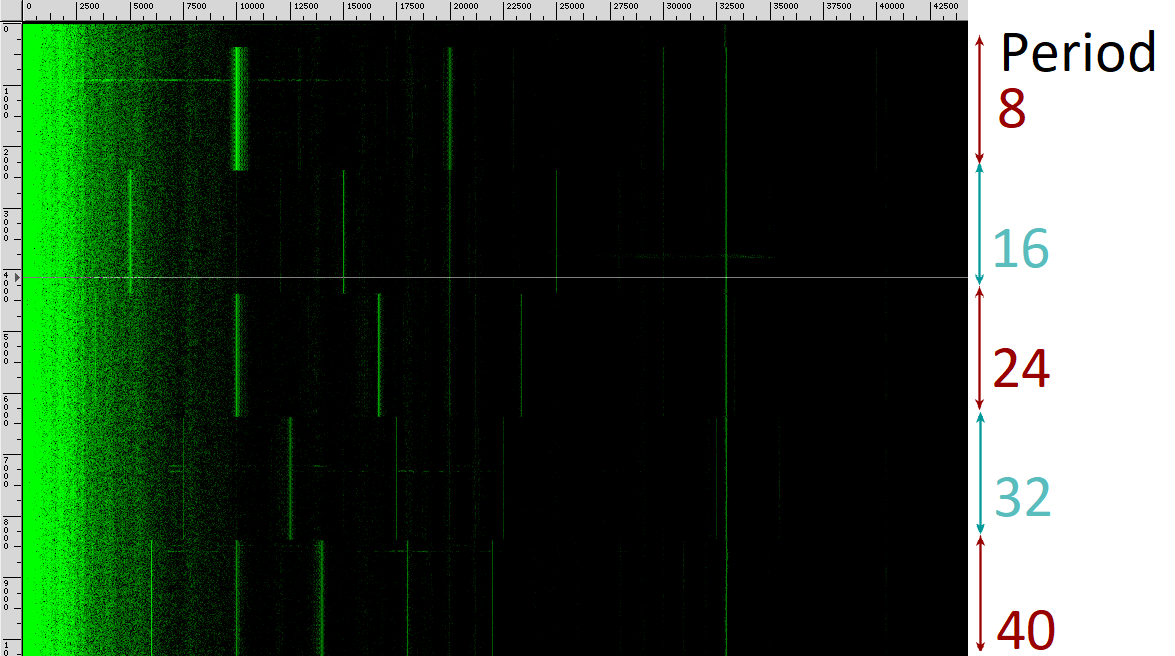}
	%COMMENT
		}
	%COMMENT
		\caption{Acoustic emanations of various target screens while displaying Zebra patterns of different periods, recorded using our setup from Section~\ref{background}.  The horizontal axis is frequency (0-43kHz) and vertical axis is time (10 seconds).}
		\label{fig:monitor-many}
		\iftwocol{}
		\vspace{-2em}
		\fi{}
\end{figure*}
}

\label{sec:acoustic-or-em}
\parhead{Acoustic or EM?} To verify that the obtained signal indeed results from an acoustic signal emanating from LCD monitors, as opposed to electromagnetic
radiation accidentally picked up by the microphone, we placed a sheet of non-conductive sound-absorbing material (e.g., thick cloth or foam) in front of the microphone. Figure~\ref{fig:monitor-single-covered} is analogous to Figure~\ref{fig:monitor-single} using the same setup, but with the microphone physically blocked with a thick piece of cloth. As can be seen in the figure, the resulting signal is severely attenuated, thus proving that the signals observed here are indeed acoustic. Additional evidence is that as microphone-to-screen distance increases, the induced signal delay matches the speed of sound rather than the speed of light (see Appendix~\ref{sec:distancevsquality}).%COMMENT
%COMMENT
\footnote{
\later\eran{SANITIZED}
Conversely, with some of the microphones used in the self-measurements experiments of Section~\ref{sec:voip}, the observed signal appears to have both acoustic and electromagnetic contributions (judging by the above methodology). This is, presumably, due to poor shielding and grounding on those cheap microphones. However, in the self-measurement setting, the precise physical channel through which the emanation propagates is inconsequential.
}

{
	\begin{figure}[t]
		\makeatletter
		\renewcommand{\p@subfigure}{\thefigure--}
		\makeatother
                        \includegraphics[width=0.55\columnwidth]{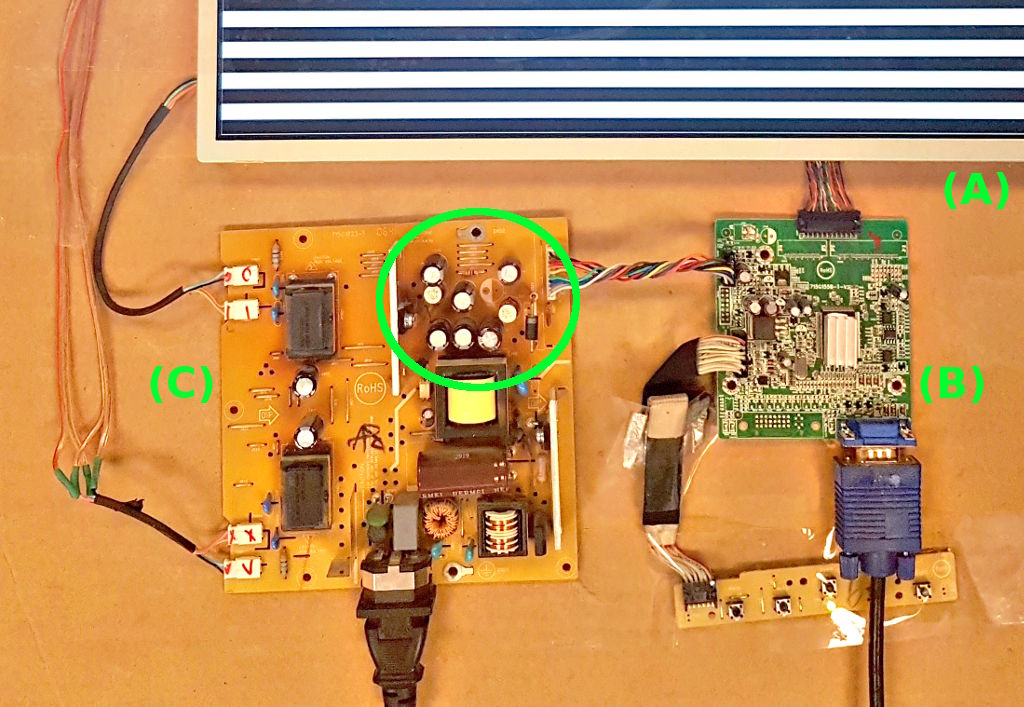}
                         \hfill
                        \includegraphics[width=0.42\columnwidth]{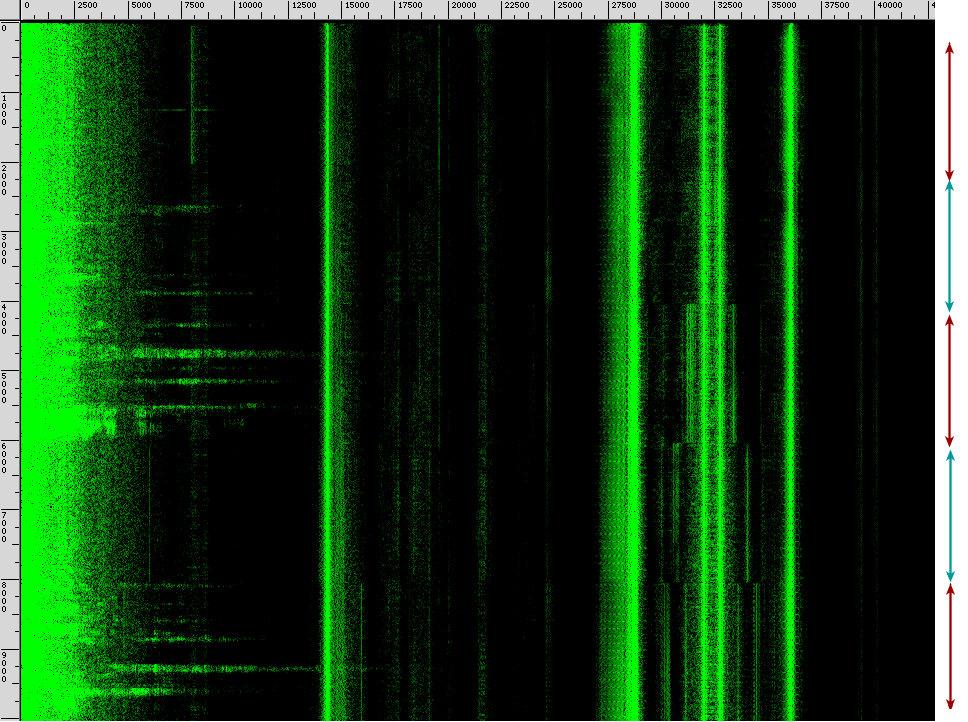}
%COMMENT
		\caption{Internal components of a ViewSonic VA903b monitor. (A) is the LCD panel, (B) is the screen's digital logic and image rendering board and, (C) is the screen's power supply board. The green circle marks the approximate source of the acoustic signal. On the right is the acoustic Zebra leakage (similarly to Figure~\ref{fig:monitor-many}) from the disassembled screen.}
		\label{fig:monitor-guts-out}
	\end{figure}
}

\parhead{Physical leakage source} Having established the acoustic nature of the leakage signal, we attempted to locate its source within the internal display electronics. To that end, we disassembled a ViewSonic VA903b LCD monitor, which has a simple, modular internal design with a moderate number of internal components. As can be seen in Figure~\ref{fig:monitor-guts-out}, in addition to the LCD panel itself (A) the ViewSonic monitor has two main boards: a digital board (B) which is responsible for implementing the monitors logic and picture rendering functions
and a power supply board (C) which provides stable voltage to the digital circuits and backlight but does not directly process the screen's content. 

In an attempt to locate the component producing the acoustic leakage, we measured the acoustic signals in various locations on both boards while displaying Zebra patterns on the LCD  panel. We localized the source of emanation to within the monitor's power supply board, in the area that is responsible for supplying power to the monitor's digital board (circled in green). We were unable to localize the source down to a single component, presumably due to acoustic reflections, diffraction and mechanical coupling in the board.%COMMENT
%COMMENT
\footnote{In an attempt to mitigate these, we also tried to lift major components away from the board and connect them by extension wires; but this changed the circuits behavior and resulting signal, and did not yield conclusive results.} %COMMENT

\parhead{Leakage Mechanism}
We conjecture that the momentary power draw, induced by the monitor's digital circuits, varies as a function of the screen content being processed in raster order. This in turn affects the electrical load on the power supply components~\cite{kjjr11introduction} that provide power to the monitor's digital board~\cite{enev2011televisions}, causing them (as in \cite{GST14acoustic-norefs}) to vibrate and emit sound. Unfortunately, the precise propagation of signal along this causal chain is difficult to precisely characterize: power signal modulation have a complex dependence on circuit topology, layout and environment; inadvertent power-to-acoustic transduction varies greatly with circuits electrical and mechanical characteristics; the different channels involve different means of acquisition and have different bandwidth and SNR constraints, necessitating different signal processing; and the choice of acoustic sensors and their placement creates additional considerations that cannot be separated from the transduction's physical nature. We thus focus on the aggregate bottom-line effect and its exploitability in the practical scenarios (e.g., webcams and phones).

%COMMENT
%COMMENT
%COMMENT
%COMMENT
%COMMENT
%COMMENT
%COMMENT
%COMMENT
%COMMENT
%COMMENT
%COMMENT
%COMMENT
%COMMENT
%COMMENT
%COMMENT
%COMMENT
%COMMENT
%COMMENT
%COMMENT
%COMMENT
%COMMENT
%COMMENT
%COMMENT
%COMMENT
%COMMENT

\parhead{Analyzing the leakage frequency} Having verified that different images produce different acoustic leakage, we now turn our attention to the connection between the displayed image and the produced leakage signal. Indeed, assume that the rendering of individual pixels produce sounds that depend on the individual pixel's color (e.g., black or white). 
Because the entire screen is redrawn approximately $r=60$ times per second, we expect that spatial-periodic changes in the displayed pixels will introduce a strong frequency component corresponding to the frequency of transitions between rendering black and white pixels. 

More specifically, a Zebra with a 16-pixel period drawn on a screen with a vertical resolution of 1050 pixels produces $1050/16=65.625$ color changes per screen redraw. Accounting for approximately 60 redraws per second, the acoustic leakage is expected to have a strong $65.625\cdot60=3937.5$ Hz frequency component; which is clearly observable in Figure~\ref{fig:monitor-single} (marked by a yellow arrow).

%COMMENT
%COMMENT
%COMMENT
%COMMENT
%COMMENT
%COMMENT
%COMMENT
%COMMENT
%COMMENT
%COMMENT
%COMMENT
%COMMENT
%COMMENT
%COMMENT
%COMMENT
%COMMENT
%COMMENT
%COMMENT
%COMMENT
%COMMENT
%COMMENT
%COMMENT
%COMMENT

\subsection{Signal analysis in the time domain}
\label{sec:timedomainanalysis}
Following our hypothesis that the intensity of pixel lines is directly manifested in the leakage signal,
we monitored the leakage of a Soyo DYLM2086 screen (see Figure~\ref{fig:balck-hole}) while displaying the \emph{Punctured Sinusoidal Zebra} image shown in 
Figure~\ref{fig:balck-hole-image}. This is a Zebra pattern with a 21-pixel period, modified in two ways. First, the intensity change between stripes is smoothed, following a sinusoidal pattern of the given period (to minimize harmonic distortions). Second, we ``punctured'' it by placing a full-width third-height black rectangle at the center of the image (to induce clearly-observable times when the sine waves disappears and reappears).
 
As can be seen in Figure~\ref{fig:balck-hole-spectrogram}, displaying the Punctured Sinusoidal Zebra causes a strong leakage at the Zebra's ``natural'' frequency of $1050/21 \cdot 60 = 3000$Hz on the resulting spectrogram. Next, plotting the\full{SANITIZED} acoustic signal in the time domain (after applying a band-pass filter at 3-4kHz) has resulted in a 3\kHz signal (matching the Zebra's frequency), where the amplitude of the middle third of the signal is considerably lower compared to the first and last third (see Figure~\ref{fig:balck-hole-baseband}). Experimenting with different sizes for the solid black area, the lower-amplitude area increases and decreases correspondingly. We conclude that the  momentary amplitude of the acoustic leakage signal is a rather accurate approximation of the brightness level of the individual pixel rows of the target monitor.

{
\begin{figure*}[t]
	\centering
\makeatletter
\renewcommand{\p@subfigure}{\thefigure--}
\makeatother
\begingroup
\centering
%COMMENT
\subfloat[Punctured Sinusoidal Zebra image with period 21]{
	\iftwocol{}
  \includegraphics[width=0.35\textwidth]{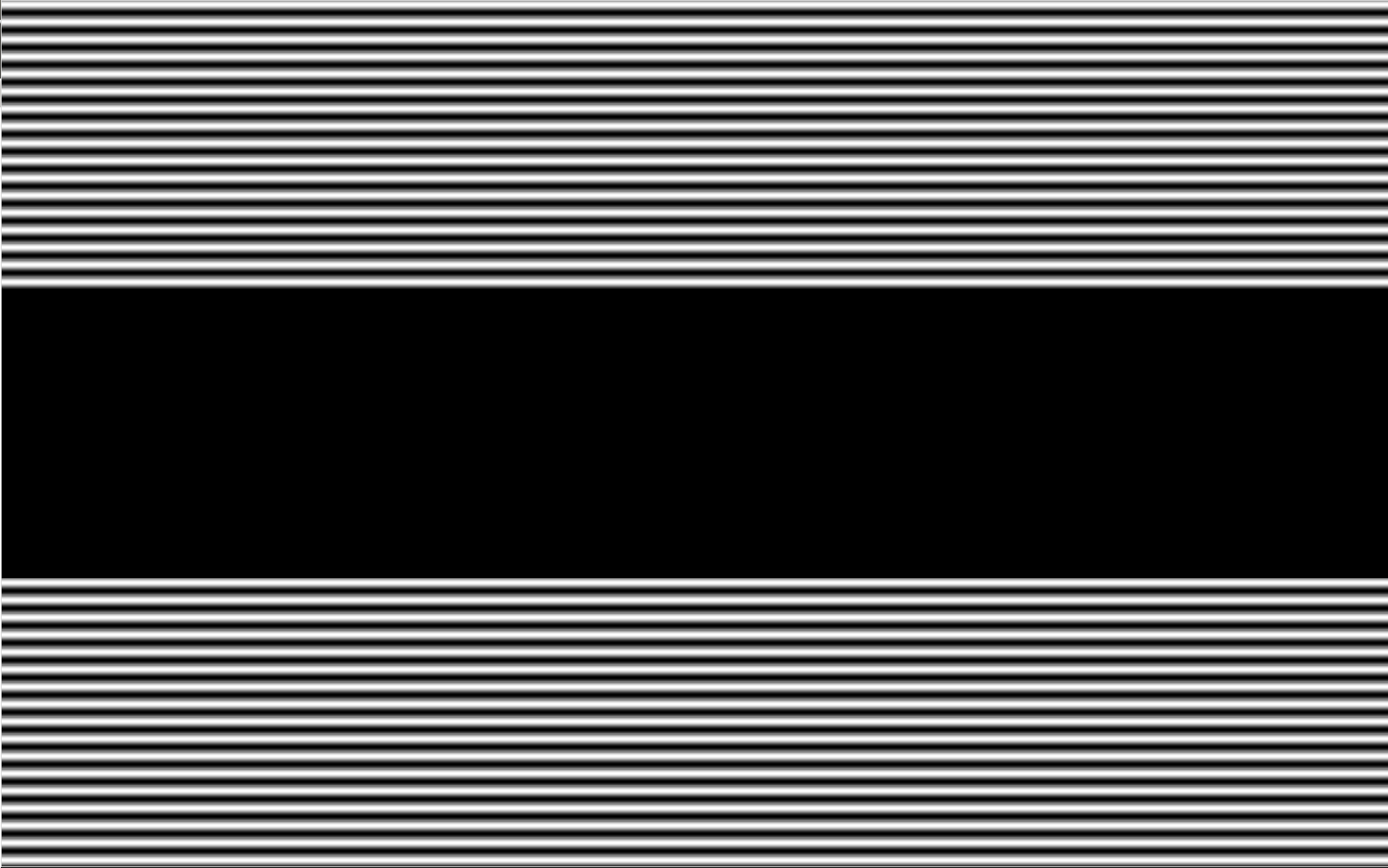}
	\else{}
  \includegraphics[width=0.45\textwidth]{art/daniels_images_final/sine_period_2_with_blackhole_vis.png}
	\fi{}
  \label{fig:balck-hole-image}
}
\hfill
\subfloat[Spectrogram (0-43 kHz, 800 ms) of the resulting acoustic signal. Notice the strong acoustic signal present at 3 kHz.]{
	\iftwocol{}
  \includegraphics[height=0.17\textheight, width=0.4\textwidth]{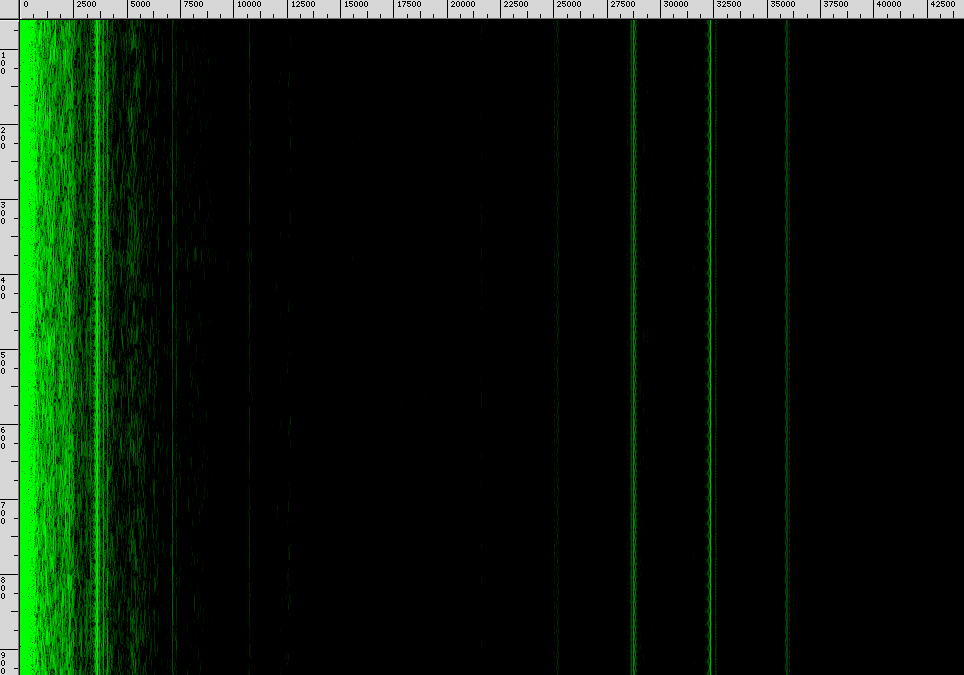}
	\else{}
  \includegraphics[width=0.45\textwidth]{art/daniels_images_final/sine_period_2_with_blackhole.png}
	\fi{}
  \label{fig:balck-hole-spectrogram}
}
\hfill
\subfloat[A segment of the acoustic signal after applying a 3-4 kHz bandpass filter. Notice the clear correlation between the amplitidue of the red signal and the brightness of pixel lines.]{
	\iftwocol{}
  \includegraphics[width=0.4\textwidth]{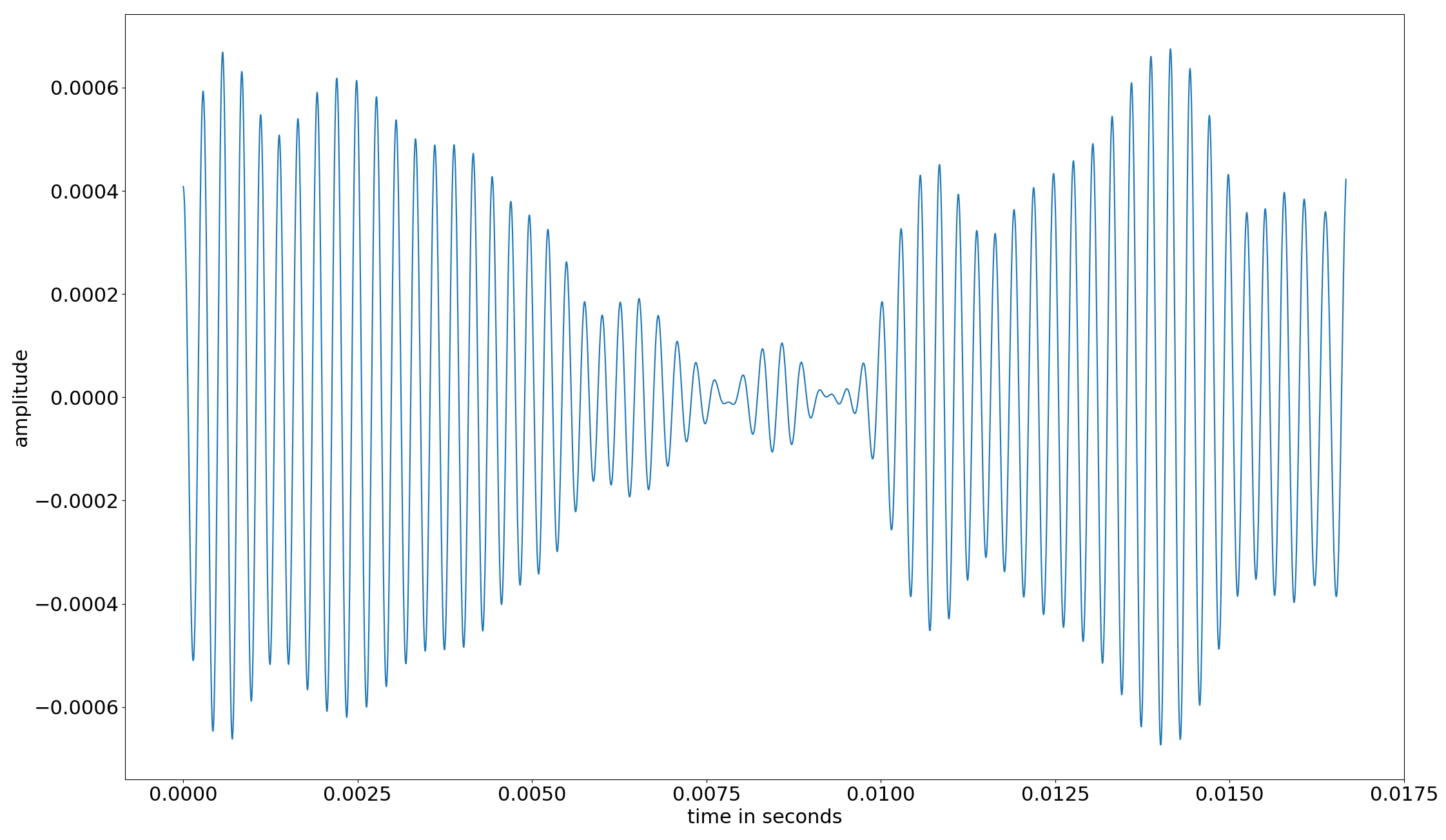}
	\else{}
  \includegraphics[width=0.45\textwidth]{art/daniels_images_final/sine_period_2_with_blackhole_time_domain.png}
	\fi{}
  \label{fig:balck-hole-baseband}
}
\hfill
\subfloat[Blue: The acoustic signal after applying a 27.5-38 kHz bandpass filter around the screen's carrier. Red: amplitude demodulation of the blue signal. Notice the inverse correlation between the amplitude of the red signal and the brightness of pixel lines.]{
	\iftwocol{}
	\includegraphics[width=0.4\textwidth]{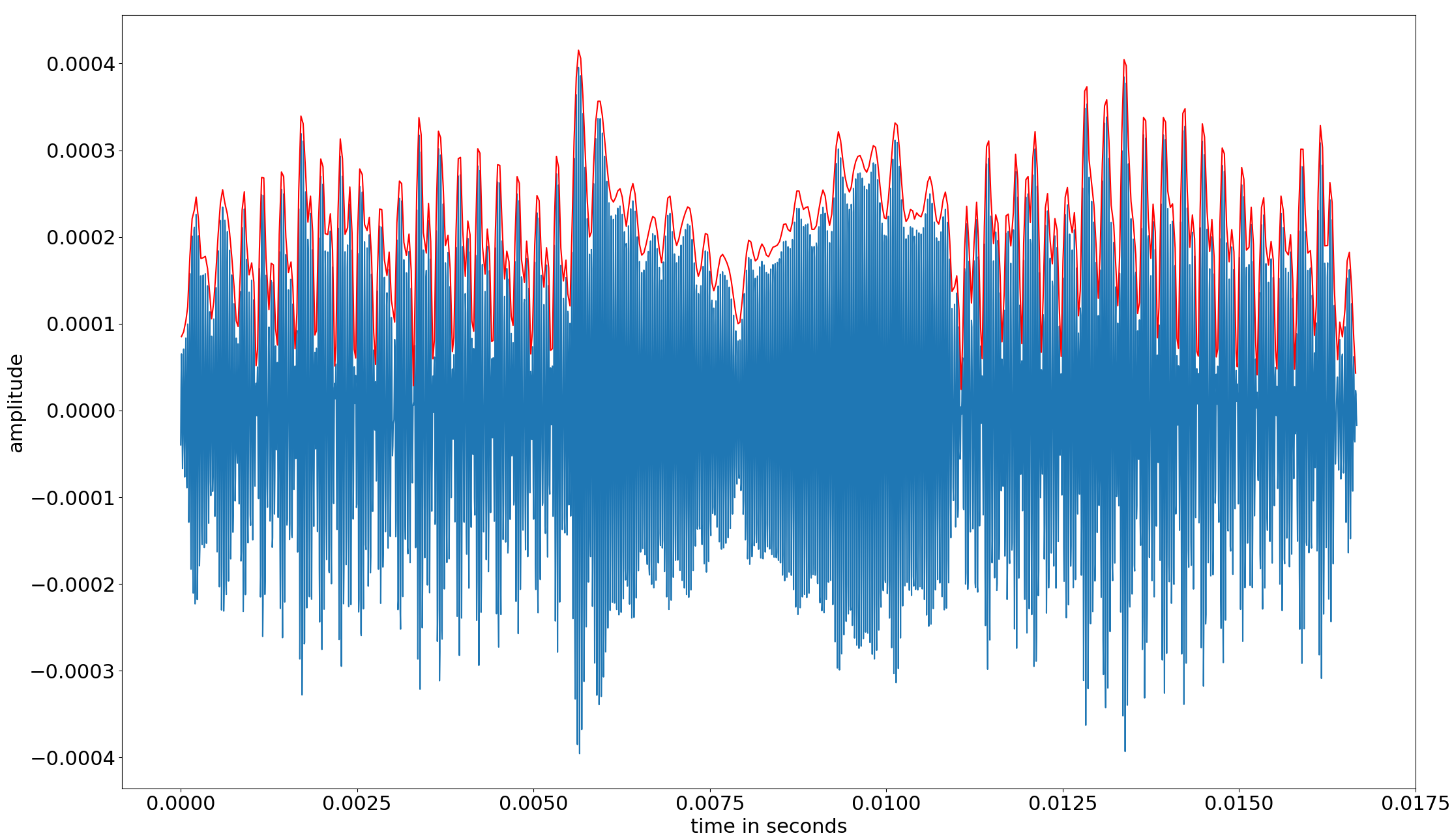}
	\else{}
	\includegraphics[width=0.45\textwidth]{art/daniels_images_final/carrier_bandpassed.png}
	\fi{}
	\label{fig:balck-hole-am}
}
%COMMENT
\endgroup
	\caption{Analyzing the acoustic leakage of a Soyo DYLM2086 screen. Recording captured using our setup from Section~\ref{background}. Figures~\ref{fig:balck-hole-baseband} and~\ref{fig:balck-hole-am} are time-synchronized using the vsync probe.}
\label{fig:balck-hole}
\vspace{-1em}
\end{figure*}
}

\subsection{Analyzing amplitude modulated signals}
\label{signalprocessing}

Observing the spectrogram in Figure~\ref{fig:monitor-single}, we notice that in addition to signal changes in the $0-22$\kHz range which correspond to different Zebra patters,  there are also changes at the $27-37$\kHz range which are again correlated to different Zebra patters. Analyzing the latter range, we notice that the signal takes form of two distinct  side lobes that mirror a central 32\kHz carrier and that increasing the period of the Zebra image being displayed leads to the carrier's side lobes coming closer together. This signal behavior at the $27-37$\kHz range is highly indicative of a different type of leakage (compared to the $0-22$\kHz range) where an amplitude modulated signal being transmitted on a central carrier. Investigating this hypothesis, in Figure~\ref{fig:balck-hole-am} we plot in the time domain the signal obtained after applying a 27.5-38 \kHz bandpass filter around the carrier and its two side lobes. As can be seen, there is a clear correlation between the color of the displayed pixel row and the amplitude of the 32\kHz carrier (blue), where dark rows corresponding to high signal amplitudes. Recovering the envelope signal (red line in Figure~\ref{fig:balck-hole-am}) using the Hilbert transform method~\cite{hahn1996hilbert} results in a direct leakage of the color of the displayed pixel row. 
%COMMENT

\begin{comment}
\begin{figure}[t]
\subfloat[The carrier signal. We applied a bandpass 27\kHz--38\kHz. The changes in signal amplitude correspond to pixel line intensity.
  \eran{SANITIZED}
  \eran{SANITIZED}
]{
  \includegraphics[width=0.5\textwidth]{art/zebra_bandpassed_with_blackhole_vis.png}
}
\caption{}
\label{ammod}
\end{figure}

\begin{figure}[t]
\subfloat[The carrier signal. We applied a bandpass 27\kHz--38\kHz. The changes in signal amplitude correspond to pixel line intensity.
  \eran{SANITIZED}
  \eran{SANITIZED}
  \eran{SANITIZED}
]{
  \includegraphics[width=0.5\textwidth]{art/zebra_bandpassed_with_blackhole.png}
}
\caption{}
\label{ammod}
\end{figure}
\end{comment}

\parhead{Leveraging the modulated signal.}
As the modulated signal is present on a carrier frequency that is much higher than the frequency of most naturally occurring acoustic signals (e.g., human speech), it can be leveraged to extract a relatively clean version of the screen's image. We now proceed to describe our method for performing acoustic image extraction. 

Because the screen refreshes approximately 60 times a second, the \textit{output trace} produced by our extraction process is a 192\kHz-sampled time series vector in the duration of one refresh cycle ($1/60\second$). Ostensibly, we simply need to sample the demodulated signal for one refresh cycle. However, as the leakage signal is not completely noise free,  we leverage the fact that the entire image is redrawn 60 times per second (thus producing a fresh sample of the leakage signal corresponding to the image at each redraw) to reduce noise by averaging. More specifically, after recording the screen's acoustic emanations for a few seconds, we bandpass filter the obtained recording around the carrier signal (allowing only the frequencies 27.5--38) and AM-demodulate it by subtracting the trace's average, computing the analytic signal using a Hilbert transform~\cite{hahn1996hilbert}, and taking its absolute value. We then divide the signal into \emph{chunks} of samples where each such chunk contains the acoustic leakage produced during a single screen refresh cycle. Finally, we perform sample-wise averaging of the chunks.

For traces acquired while the vsync probe is attached (see Section~\ref{background}), we can easily apply this approach: after demodulating, we chop the signal into chunks according to probe's signal, so they correspond exactly with refresh periods, and average. Figure~\ref{fig:triggered-blackhole} illustrates this.

\parhead{Modulated signal quality.}
We observe that while individual chunks exhibit some noise-related variations, they do tightly follow a similar content-dependent pattern, with an average Pearson correlation coefficient of 0.957 between a chunk and the average of all chunks. We used this correlation test to perform a systematic investigation of the effect of distance on the leakage signal (see Appendix~\ref{sec:distancevsquality}). We found that by this metric, the content-dependent leakage is observable when recording at a distance of up to 3 meters (by a bare microphone without a parabolic dish).

\begin{figure}[t]
\iftwocol{}
\vspace{-2em}
\fi{}
\centering\includegraphics[width=0.7\columnwidth]{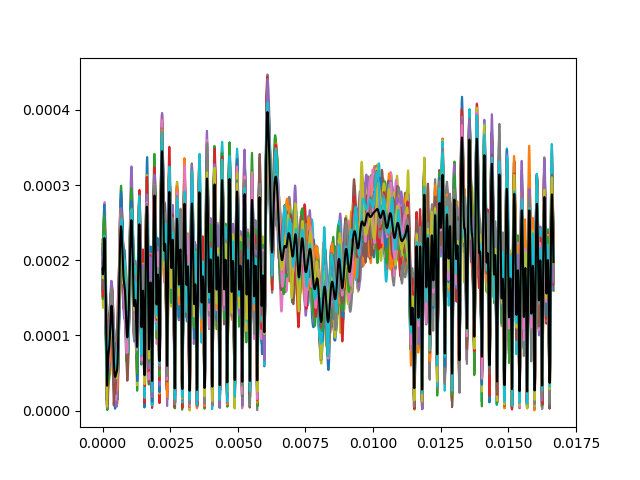}
\vspace{-1em}
\caption{
\label{fig:triggered-blackhole}
	One-period modulated signal chunks (in color) and their average (in black), of the ``black-hole'' screen trace (see Figure~\ref{fig:balck-hole-image}).
}
\iftwocol{}
\vspace{-1em}
\fi{}
\end{figure}

\begin{comment}
\begin{figure}[t]
%COMMENT
  \includegraphics[width=0.5\textwidth]{art/demod.png}
%COMMENT
%COMMENT
\caption{
  AM demodulation, using the absolute value of the Hilbert transform
  \eran{SANITIZED}
}
\label{demodFig}
\end{figure}
\end{comment}

%COMMENT

\subsubsection{Chunking challenges}
\label{sec:chunkchall}
The vsync probe we use above is not available to the attacker, who must nevertheless divide the signal accurately into chunks whose phase within a refresh period is almost identical. Next, we discuss the challenges in this, and our approach in addressing them.

\parhead{Challenge 1: drift.}
Unfortunately, using exactly 60\Hz is usually erroneous by up to 0.2\Hz. The refresh rate used has to be extremely accurate: in our sample rate (192\kSps) and around a 60\Hz refresh rate, an error of about 0.002\Hz would cause a drift of one sample per ten chunks. Even such a small error would introduce destructive interference in averaging hundreds of chunks. The attacker thus has to very accurately measure the refresh rate. Empirically, even for the exact same configuration, the refresh rate has slight changes of up to 0.02\kHz.
%COMMENT

\parhead{Challenge 2: jitter.}
Using the vsync probe (see Section~\ref{background}), we examined refresh cycles. We observed signal jitter that caused destructive interference even when averaging with the accurate refresh rate. The jitter creates timing abnormalities that are not averaged out over a few seconds. For our Soyo monitor, every few dozen refreshes, there is an \textit{abnormal cycle}, whose duration is abnormal. The abnormalities are erratic: cycles can be either down to 10\% shorter than  $1/60\second$ long, or as long as  $1/30\second$. \done\roei{SANITIZED} \done\daniel{SANITIZED}Even with the accurate average refresh rate, drift still occurs: the average refresh rate is actually far from the refresh rate for non-abnormal refresh cycles. Effectively, every few hundred cycles, the refresh period gets phase-shifted by some unknown value.

\parhead{Naive solution: correlation.}
A natural approach \later\roei{SANITIZED}is first approximating or measuring the refresh rate, segmenting the trace into chunks corresponding to refresh cycle duration, and use correlation with one arbitrarily chosen ``master'' chunk to phase-align the traces, i.e., trace phases are aligned such that correlation is the highest. Then, outlier rejection can be performed to mitigate jitter and drift in still-misaligned chunks. We found the variants of this approach to underperform even when the refresh rate used is extremely accurate (see Appendix~\ref{sec:algocompare}). Presumably, this is because they all allow for very lenient alignment of the chunks to the master chunk,  i.e., all chunks are rotated so that they correlate best with the master chunk, which contains a noisy version of the signal of interest. Averaging is thus likely to increase the noise in the master chunk. Another problem is arbitrarily setting the master chunk, which can produce an abnormally noisy chunk. The approach is also prohibitively slow.

\later\eran{SANITIZED}

\subsubsection{Our denoising approach}
\label{denoising}
We need a chunking algorithm that is robust to jitter and small drift, i.e., every chunk should correspond to a refresh cycle at a specific phase (identical for all chunks). Denote the sample rate $f_{s}$. We empirically observe that, except for abnormal cycles, refresh times are relatively stable in the following sense: for each of our recordings there exists an integer $W\approx \frac{f_{s}}{r}$ such that cycles are either $W$ or $W+1$ samples long, implying that the actual refresh cycle is between $\frac{f_{s}}{W}$ and $\frac{f_{s}}{W+1}$. Moreover, $W$ is always one of two possible consecutive numbers: $S,S-1$. Non-abnormal cycle sizes are therefore always in ${S-1, S, S+1}$. Another empirical observation is that Pearson correlation values can be used to heuristically distinguish between pairs of chunks with the same cycle phase and ones with different phases: same-phase chunks are typically above some threshold value, whereas different-phase chunks are typically below it (especially if the phase difference is more than a few samples). Our chunking algorithm is parametrized by $S$, by a small integer $d$, and a ``correlation threshold'', a real number $T$.

\parhead{High-level overview}
The algorithm starts from the first sample in the signal, and iteratively finds the next chunk start location using Pearson correlation with a reference chunk, assuming that each chunk size is in $G=\{S-d, ..., S+d\}$. The range $d$ is very small: in our experiments it is usually 1, and never more than 3. This is designed to make small adaptive changes in chunk sizes to account for the minor drift introduced by the refresh-rate approximation and, sometimes, misalignment made in the previous round. When correlation of the reference chunk with the next chunk drops below $T$, we enter a ``sync'' mode where the purpose is to re-synchronize the chunk phase, again using correlation. If the algorithm enters sync mode for more than 15\% of the signal chunks, or if the algorithm does not exit sync mode after three iterations, it quits with an error. After averaging, we rotationally shift the resultant array such that the highest value is the first.\footnote{This results in consistent trace phase alignment since, as we found, the highest-value sample corresponds to one of a few refresh period phases.} Appendix~\ref{chunkingalgo} describes our algorithm in detail.

\parhead{Choosing parameter values.}
$d$ is chosen to be very small to ensure we are not increasing the noise by maximizing correlation (see advantages below). We used 3 for the keyboard snooping and cross-screen attacks, and 1 for the other ones. $T$ should be the threshold that differentiates best between out-of-phase chunk pairs and in-phase chunk pairs. For clean signals such as the close-range attack, we chose 0.9. For noisier signals we chose 0.4-0.8, accepting some ``false'' entries into sync mode. We measured $S=3206$ to $S=3202$ empirically using the vsync probe. The attacker, however, does not need to attach a probe to the victim computer: since the range of possible values is small, the attacker can apply brute force guessing and, using our algorithm, choose the value of $S$ that maximizes average chunk correlation with the master chunk and minimizes sync mode iterations. This approach
%COMMENT
accurately finds the correct $S$ value using under a minute of victim screen recordings. In our ``cross-screen'' experiments described throughout the paper (e.g. in Section~\ref{sec:crossscreenmany}), we performed attacks without attaching a vsync probe to the victim screens.
%COMMENT

%COMMENT

\parhead{Advantages of this approach.}
First, it is less likely to augment the noise introduced by the master chunk. This is because of its strict limitation: the vast majority of chunks found have to be consecutive and with a very tight size constraint. Only the first chunk in each sequence is leniently aligned such that it correlates best with the master chunk. 

Second, the master chunk is not arbitrarily found, but is a chunk that correlates well with at least its consecutive chunk. An abnormally noisy chunk is less likely to be selected.

Third, the algorithm is not parametrized by the exact refresh rate. Instead, it can use an approximation. Not depending on the exact refresh rate is crucial, since it slightly changes with time, so the attacker cannot be expected to track it.

Finally, this algorithm is much faster than the rotational-shift-based baseline. The complexity of both methods is dominated by the required Pearson correlation computations, an operation with complexity $O\left(S\right)$. The baseline's approach computes the Pearson correlation and performs this $C\times S$ times, where $C$ is the number of chunks. For a 5\second trace at 192\kHz sample rate, our implementation of this approach takes 166\second on an Intel Xeon E5-2609 v4. While in the worst case ($6000\times 0.15 \times C$ correlation operations in sync mode, and another $2\times d\times 0.85\times C$ correlation operations in normal mode), our algorithm performs only about three times fewer correlation computations as the baseline; in practice it is over two orders of magnitude faster on average. Even for recordings captured using a parabolic mic, where the signal is noisy and sync mode is entered relatively often, the average processing time is less than 2\second for 5\second recordings.

Figure~\ref{fig:algorithm-output} shows the output trace for the Punctured Sinusoidal Zebra image. Appendix~\ref{sec:algocompare} demonstrates how our approach produces less noisy output traces than a natural baseline.

\begin{figure}[t]
\iftwocol{}
\fi{}
\centering\includegraphics[width=0.65\columnwidth]{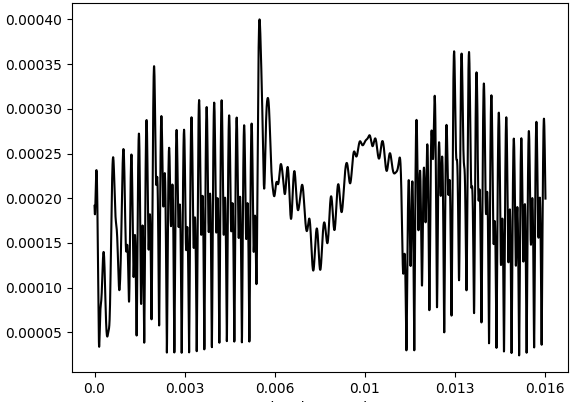}
%COMMENT
\caption{
\label{fig:algorithm-output}
	The output trace of the image of~\ref{fig:balck-hole-image}, using $T=0.9, S=3206, d=3$. This is visually indistinguishable from the product of chunking and averaging using the vsync probe (see Figure~\ref{fig:triggered-blackhole}).
}
\iftwocol{}
\vspace{-1em}
\fi{}
\end{figure}

\subsection{Cross-screen signal similarities}
\label{sec:crossscreenzebra}

From a remote attacker's perspective, the existence of leakage is not enough. If the leakage does not behave consistently across screens, it will be hard for an attacker without physical access to the victim's screen to learn how to process its emitted signal. This is particularly problematic for an attacker who trains machine learning models on the leakage signal, because they might overfit to screen-specific attributes. We now show that the relationship between content and leakage is largely predictable and similar across screens, even when the signal is recorded through different setups and in different environments.

\parhead{Visually observing similarities.}
We used two Dell 2208WFPt screens. We also used two setups: our setup from Section~\ref{background}, the \textit{close-range} setup, and one where the microphone is placed 2 meters from the screen in a parabolic dish. See Figure~\ref{fig:parabola} for an example of such a setup. Both setups were set in different offices (with different environment noises). In each setup for each screen, we displayed a Punctured Zebra (a Zebra overlayed by a full-width third-height black rectangle at its center) and recorded a 3\second trace. We applied our preprocessing procedure from Section~\ref{signalprocessing} and proceeded to visualize traces of identical content, of the same screen, from different screens, and from different screens and different recording settings. Figure~\ref{fig:cszebra} shows the result.

First, we compare Figure~\ref{fig:blacksame}, displaying two traces of the same Punctured Zebra on the same screen, and Figure~\ref{fig:blackdifferent}, displaying traces of the same pattern on different screens. Clearly, the correspondence of signal amplitude and pixel intensity is similar on both screens. Next, we notice that the  traces do display some minor screen-specific traits, making traces collected from the same screen (Figure~\ref{fig:blacksame}) to be more similar than ones collected from different screens (Figure~\ref{fig:blackdifferent}).
%COMMENT

Second, note that even two signals taken from different screens \emph{and} using different setups (Figure~\ref{fig:blackdifferentdist}) display obvious similarity: the at-distance signal is weaker, and yet has the same characteristics, with the on-screen Black Hole clearly inducing a flat line in the middle portion of both, and the zebra-like pixel color alternations manifesting as a sine wave to the right and left of the flat line. Below, we show this quantitatively.

\begin{figure*}[t]
		\vspace{-3em}
\centering
\begingroup
\captionsetup[subfigure]{width=0.31\textwidth}
	\subfloat[Different signals of the same content of the same screen, taken using the close-range setup. \vspace{-2em}]{
\includegraphics[width=0.31\textwidth]{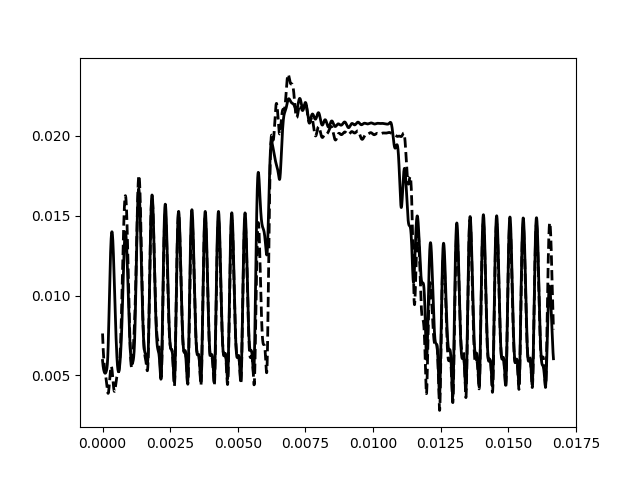}
	\label{fig:blacksame}
}
\hfill
\subfloat[Signals of two different screens, taken using the standard setup.]{
\includegraphics[width=0.31\textwidth]{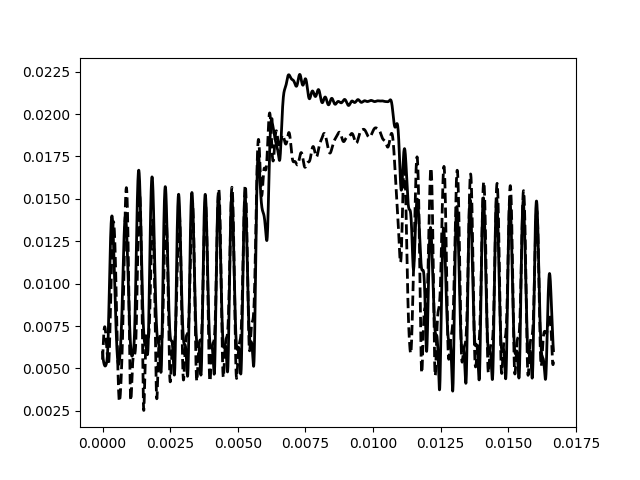}
	\label{fig:blackdifferent}
}
\hfill
\subfloat[Signals of different screens, one (solid) taken in close-range and one (dashed) taken from a distance using the parabolic dish.]{
		\vspace{-3em}
\includegraphics[width=0.31\textwidth]{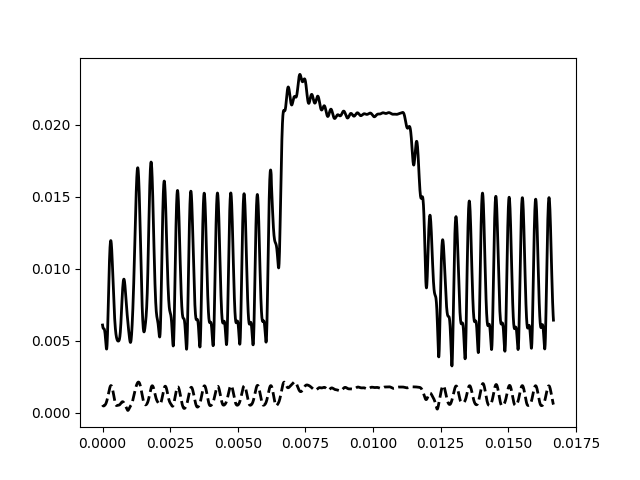}
	\label{fig:blackdifferentdist}
}
\endgroup
\caption{
\label{fig:cszebra}
	Signals recorded while screens were showing a Punctured Zebra (with 16-pixel period and black middle), after our preprocessing procedure from Section~\ref{signalprocessing} and correlation-based phase alignment.
}
\vspace{-1em}
\end{figure*}

\parhead{Quantifying similarity.}
We then applied a correlation test, which showed that a relatively strong correlation exists between signals of different screens showing the same content, even when they are recorded using different setups.
We recorded multiple traces for zebras with 8, 16, 24, 32, and 40-pixel periods, as well as, again, a Punctured Zebra with period 16 (as above). Alternating these patterns in a round-robin fashion, we captured 10 3\second traces in the close-range setting for one Dell 2208WFPt screen, and at a distance and using the parabolic dish for another. We applied our signal processing procedure to all traces. Next, we performed the following for each of the two screens: first, we randomly chose a trace for each pattern from this screen's traces. Then, for every trace from the other screen, we computed its maximal Pearson correlation over all rotational shifts, with each of the 6 chosen traces. In 100\% of the cases, the trace of the matching pattern was the one displaying the highest correlation value. We thus conclude that screen content is the dominating factor in the screen's leakage signal, while screen-specific effects can be considered as negligible.

\section{Attack vectors}
\label{attackvectors}

We consider four attack vectors. For each, we specify our experimental setup, demonstrate that leakage exists and, in following sections, evaluate screen content detection attacks.

\later\roei{SANITIZED}
%COMMENT
%COMMENT
\subsection{Close-range and at-distance attacks}
\label{sec:closevsfar}
In this setting, the attacker uses a high-end microphone to extract the image displayed on the screen. While a relatively strong attack model compared to other models considered in this paper, it is most effective for estimating the extent of acoustic information leakage from various attack ranges.

\parhead{Experimental setup}
We used a setup similar to the one in Section~\ref{background}. For the distance attack, we mounted the microphone in a parabolic dish, placed about 10 meters away from the screen. See Figure~\ref{fig:parabola} for an example of such a setup. To simulate a realistic scenario, recordings were taken in an office environment, with some noise from nearby equipment and people occasionally talking in the proximity of the microphone. We do not expect speech to interfere, as most of the leakage frequencies are well-above speech frequencies.
%COMMENT

\iftwocol{}
\begin{figure}[t]
\vspace{-1em}
\else{}
\begin{figure*}[t]
\fi{}
	\makeatletter
	\renewcommand{\p@subfigure}{\thefigure--}
	\makeatother
	\centering
        \iftwocol
          \includegraphics[width=0.5\textwidth]{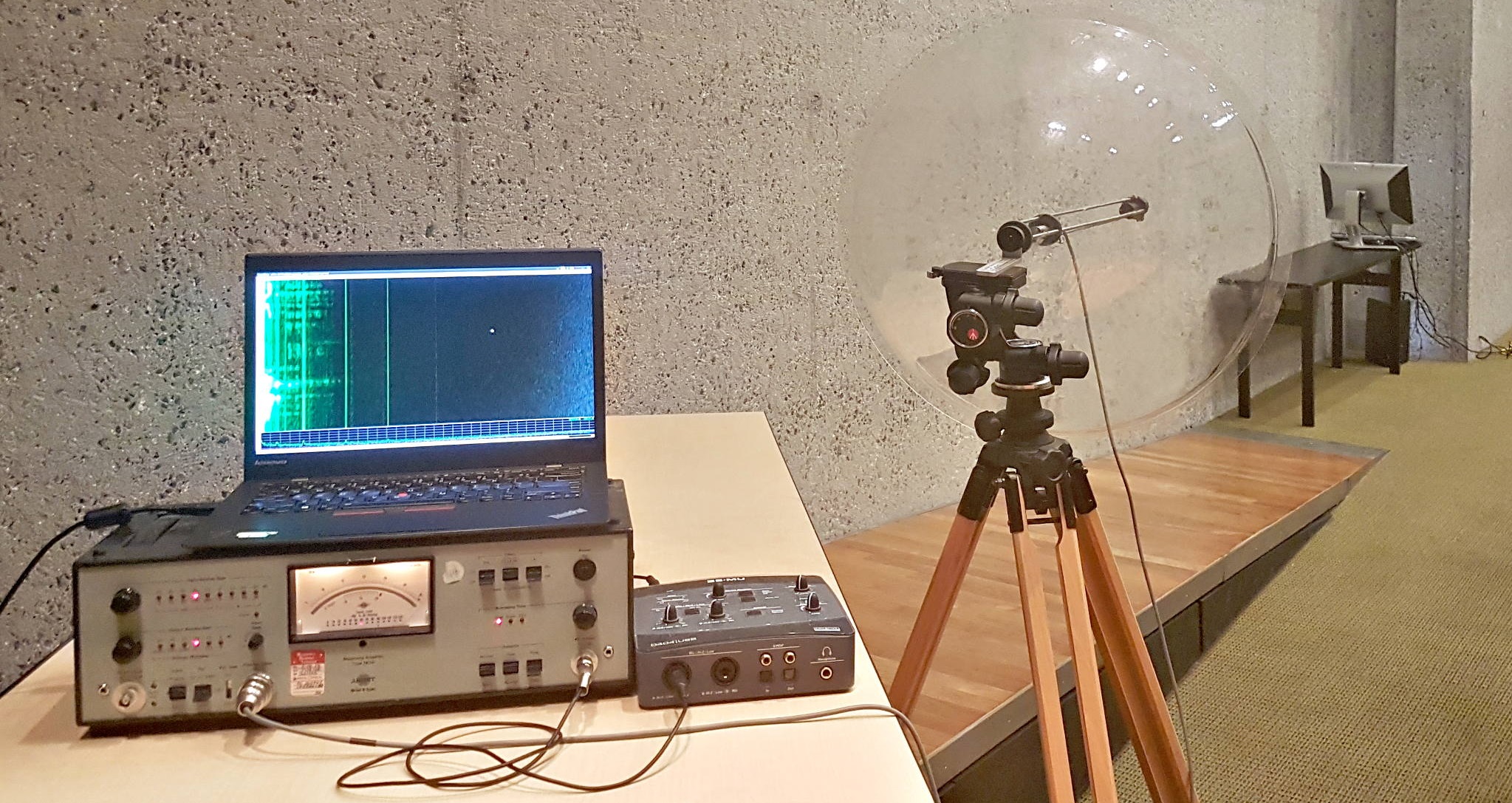}
        \else
          \includegraphics[width=0.7\textwidth]{art/attack_vectors/parabolic-setup_w2048.jpg}
        \fi
%COMMENT
	\caption{Extracting an image from a screen using a parabolic microphone at a distance of approximately 5 meters. \later\daniel{SANITIZED}}
	\label{fig:parabola}
	%COMMENT
	\vspace{-1em}
	%COMMENT
\iftwocol{}
\end{figure}
\else{}
\end{figure*}
\fi{}

%COMMENT
%COMMENT
%COMMENT

\subsection{Phone attack}
\label{sec:phoneattack}
We consider a commodity phone placed in proximity to a screen, recording its acoustic emanations using the phone's built-in microphone (see Figure~\ref{phones}).
This vector is useful in a host of scenarios, such as spying on a turned-away screen (e.g., in a business meeting) using a personal phone. The attack can also be conducted remotely by an app, without the phone owner's knowledge, because many mobile apps have permission to record audio at any time~\cite{schuster2016droiddisintegrator, felt2012android}.
%COMMENT
%COMMENT
 \later\roei{SANITIZED}

 \done\roei{SANITIZED}
\parhead{Experimental setup}
For this attack setup, we used an LG V20 phone (which supports a 192\kHz sample rate\done\roei{SANITIZED} and, empirically, exhibits good sensitivity to sound even beyond 40\kHz) directed at a screen and recording using its built-in microphone. We used the Amazing MP3 Recorder app~\cite{amazingmp3recorder}, which supports high sample rates and offers an interface we used for automating trace collection en masse: a ``start recording'' and ``stop recording'' are exported by this app and can be used by \emph{any} other app on the device through Android's inter-app communication mechanisms.~\footnote{
	The app exposes interfaces (``broadcast receivers'') that can be used by any app, regardless of its permissions, to record audio. This demonstrates the commonplace nature of mobile based audio capturing adversaries.} In particular, we used the Android Debug Bridge (ADB) interface for invoking recordings. Here, too, recordings were taken in an office environment with some environmental noise and human speech.

\parhead{Demonstrating leakage.}
We found that the phone, recording at 192\kSps sample rate, can capture the leakage extremely well. We performed an  experiment similar to those in Section~\ref{exploringleakage} twice: once with the phone directed at the back of the screen (simulating a physical attack in, e.g., a business meeting), and once with the phone naturally positioned on a table near the screen (simulating a remote attacker, e.g., an app). In both cases, the measured screen was a Soyo DYLM2086 screen. Figure~\ref{phones} shows the resulting spectrograms, containing the expected leakage pattern. While for the naturally positioned phone the signal is attenuated, it is still clearly visible on the spectrogram. The leakage signal for the directed position is dominant and is almost comparable to the leakage samples recorded using high-end equipment.

{
\begin{figure*}[t]
%COMMENT
\makeatletter
\renewcommand{\p@subfigure}{\thefigure--}
\makeatother
\subfloat[Phone microphone directed at the back of the screen.]{
  \includegraphics[height=0.170\textwidth]{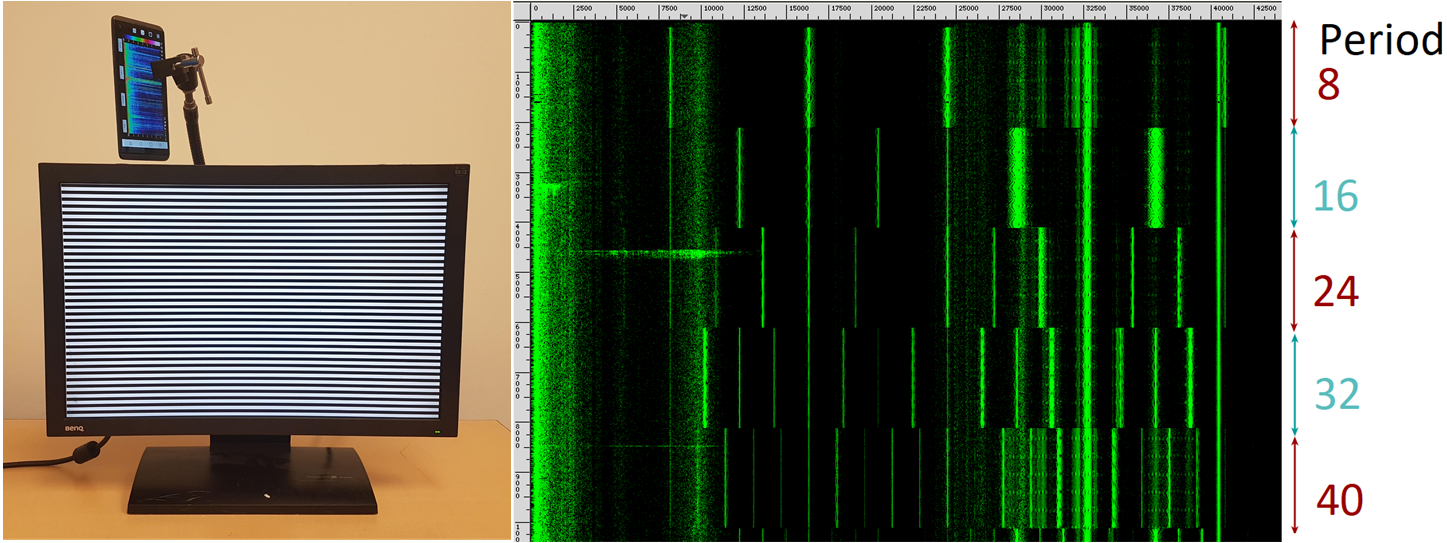}
  \label{unnaturalPhone}
}
\hfill
\subfloat[Phone placed on a desk by the screen.]{
  \includegraphics[height=0.170\textwidth]{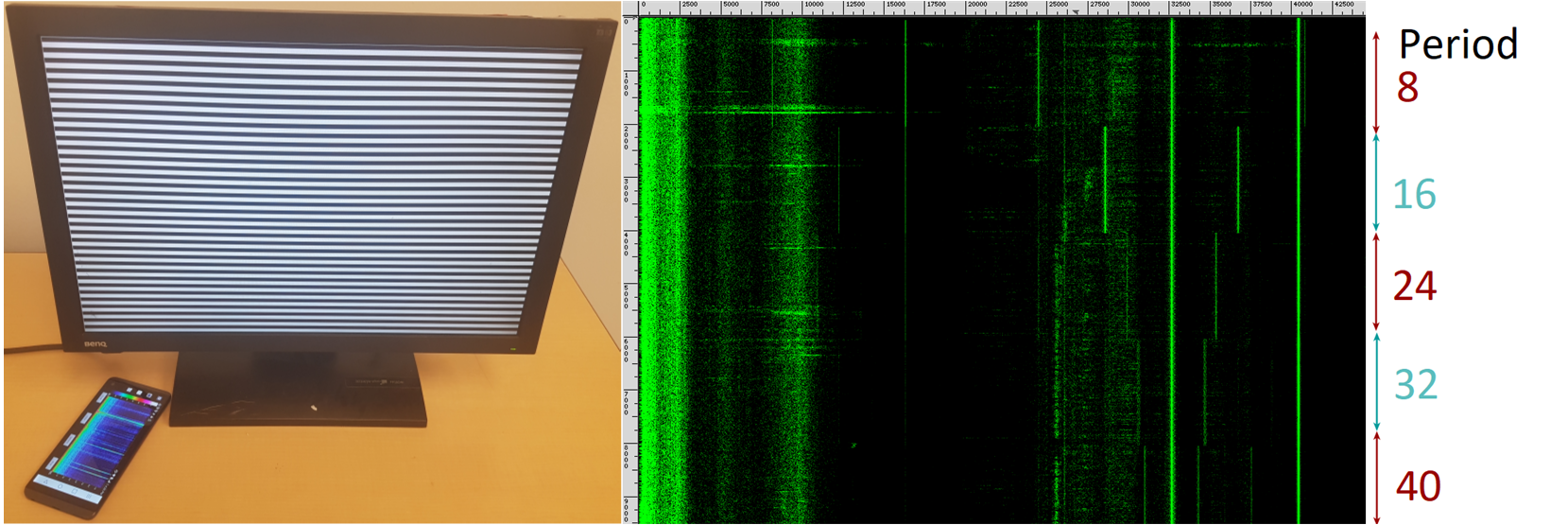}
  \label{naturalPhone}
}
%COMMENT
\caption{Acoustic emanations (0-43\kHz, 10sec) of a BenQ q20ws screen while displaying Zebra patterns of different periods, recorded using an LG V20 smartphone.}
\label{phones}
\iftwocol{}
\vspace{-2em}
\fi{}
\end{figure*}
}

\parhead{Evaluation.} We evaluate attacks using this vector: on-screen keyboard snooping is evaluated in Section~\ref{keyboardattack} and a website distinguishing attack is evaluated in Section~\ref{websitedistinguish}.

\subsection{VoIP attacker}\label{sec:voip}
We also consider a remote adversary who is conversing with the target over the Internet, using a Voice over IP (VoIP) or videoconferencing service. As long as the webcam (if any) is not pointing at the screen (or at its reflection of some object), the target would assume that remote participants in the call cannot glean information about the local screen's content. However, the adversary receives an audio feed from the target's microphone, which is usually located close to the screen (in order to maintain eye contact during video calls), or even embedded into the screen itself. This microphone picks up the screen's acoustic emanations, the VoIP service relays it digitally, and the remote attacker can (as we will show) analyze this signal to extract the screen's content.

\parhead{Experimental setup.} 
Empirically demonstrating this, we obtained the screen's acoustic emanations by recording the audio sent to a remote party during a Hangouts call, captured using the victim's own microphone (built into a commodity webcam). Here, too, recordings were taken in an office environment with some environmental noise and human speech. 

More specifically, to simulate the victim, we used a PC running Windows 10, connected to a Dell 22" LCD monitor (model 2208 WFPt) screen, and a Microsoft LifeCam Studio webcam. The camera was placed naturally by the screen, similarly to Figure~\ref{hangouts}.
To simulate the attacker, we used a second PC running Ubuntu 16.04. We set up a Hangouts connection between the attacker and victim, both running Hangouts over a Firefox browser.
At the attacker end, we redirected all sound output from the soundcard into a loopback device. We could then use \texttt{arecord} on the device to capture sound originating at the victim end of the call.
%COMMENT

%COMMENT

\parhead{Observing the leakage.}
We again performed measurements by acquiring traces while displaying alternating Zebra patterns, similar to those in Section~\ref{exploringleakage}, but using various webcams and screens. Figure~\ref{fig:webCams} summarizes our findings.

We discovered that, first, commodity webcams and microphones can capture leakage. Second, natural and expected positioning of cameras can be sufficient. In fact, even the built-in microphones in some screens (e.g., Apple LED Cinema 27-inch, see Figure~\ref{apple}) can be sufficient. Third, the leakage is present (and prominent) even when the audio is transmitted through a Hangouts call; see Figure~\ref{hangouts}.

%COMMENT

{
\begin{figure*}[t]
%COMMENT
\makeatletter
\renewcommand{\p@subfigure}{\thefigure--}
\makeatother
	\iftwocol{}
	\fi{}
	\captionsetup[subfigure]{width=\textwidth}
	\subfloat[Microsoft LifeCam Studio webcam positioned below a LED-backlit Dell U2713H screen, recorded through a Hangouts call. \done\eran{SANITIZED}
	The frequency axis range is 0-13\kHz, due to the lower sampling rate used by Hangouts.]{
		\iftwocol{}
  \includegraphics[height=0.150\textwidth]{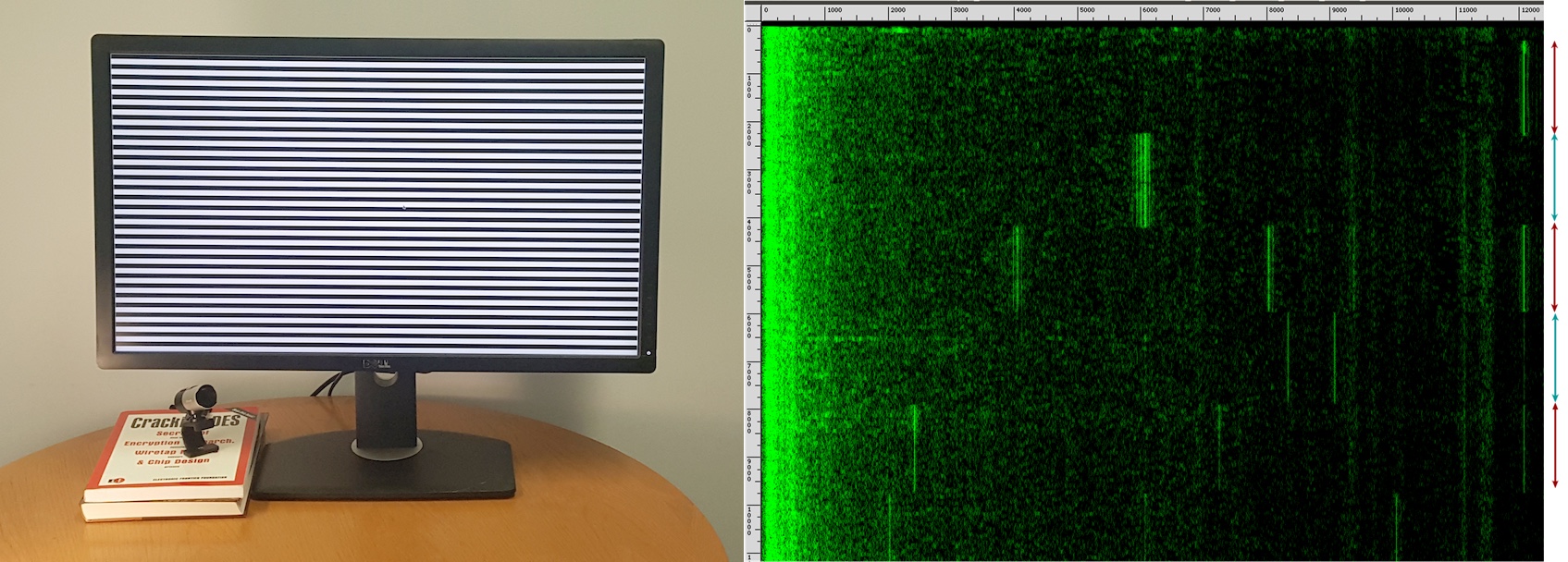}
		\else{}
  \includegraphics[width=\textwidth]{art/attack_vectors/lifecam-dell-setup-zrattle-arrowannot_new.jpg}
	\fi{}
  \label{hangouts}
}
\hfill
\iftwocol{}
\else{}
\begingroup
\captionsetup[subfigure]{width=0.45\textwidth}
\fi{}
\subfloat[Self-measurement: an embedded microphone in an LED-backlit Apple LED Cinema 27-inch screen.
  \eran{SANITIZED}
  ]{
		\iftwocol{}
  \includegraphics[height=0.150\textwidth]{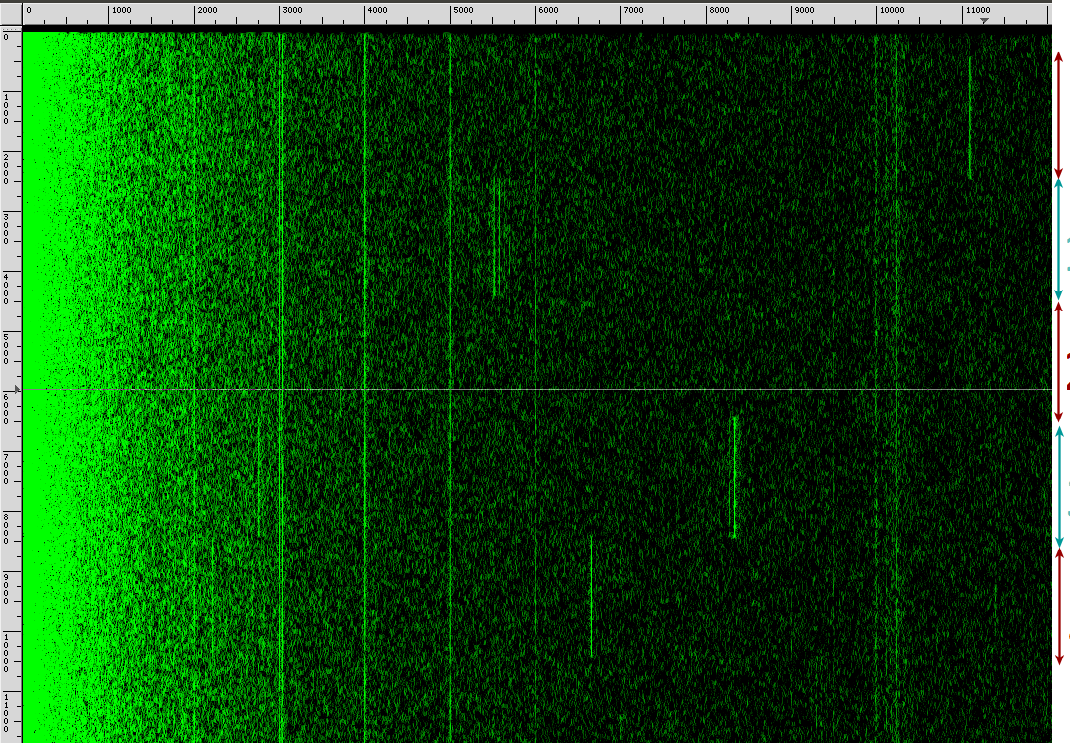}
		\else{}
  \includegraphics[height=0.31\textwidth]{art/attack_vectors/apple-arrowannot.png}
	\fi{}
  \label{apple}
}
\iftwocol{}
\else{}
\endgroup
\fi{}
\hfill
\iftwocol{}
\else{}
\begingroup
\captionsetup[subfigure]{width=0.5\textwidth, justification=raggedright}
\fi
\subfloat[Logitech C910 webcam positioned on top of a 20`` Soyo screen.]{
		\iftwocol{}
  \includegraphics[height=0.150\textwidth]{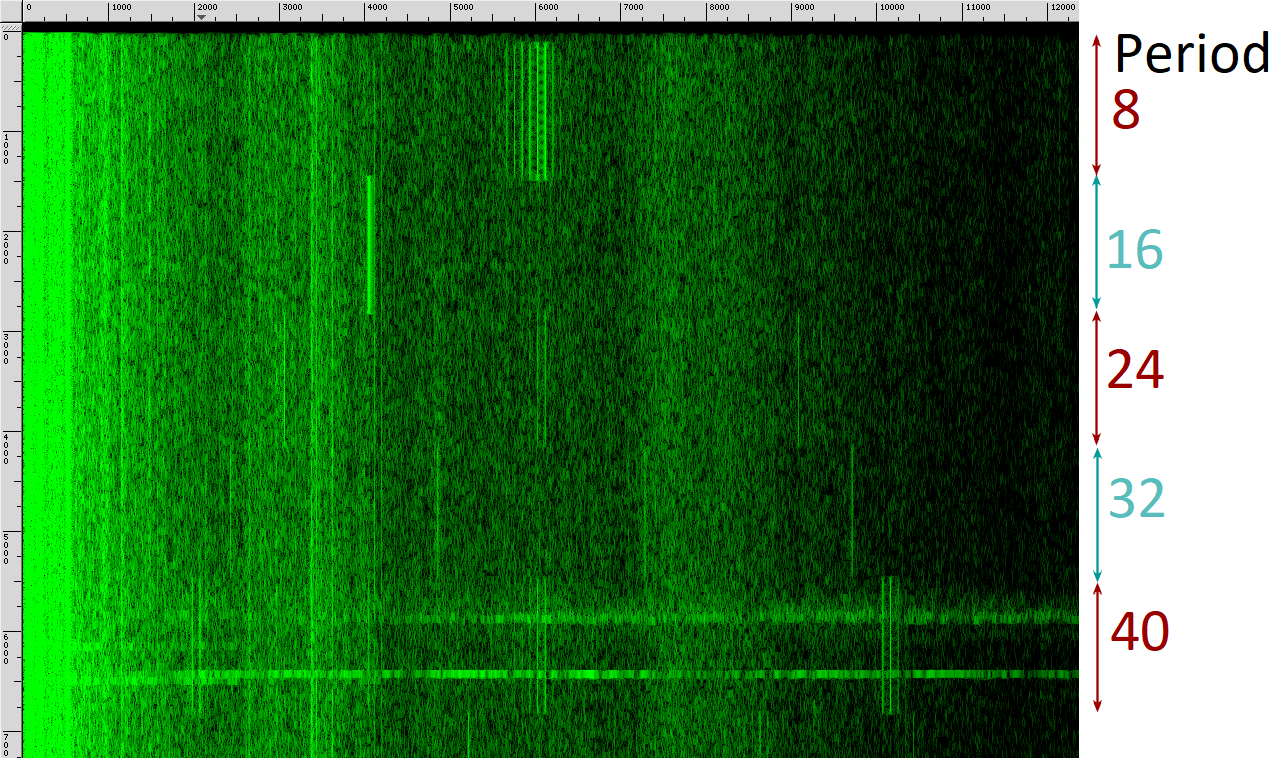}
		\else{}
  \includegraphics[height=0.31\textwidth]{art/attack_vectors/soyo_c910.png}
	\fi{}
  \label{c910}
}
\iftwocol{}
\else{}
\endgroup
\fi
\hfill
%COMMENT
\caption{Time-frequency spectrum of alternating Zebra frequencies measured through various naturally positioned commodity devices. The spectrograms of all combinations indicate similar acoustic leakage patterns.}
\label{fig:webCams}
\iftwocol{}
\vspace{-1em}
\fi{}
\end{figure*}
}

\parhead{Attack evaluation.} See Section~\ref{codecattack}.

\subsection{Virtual assistant / ``smart speaker'' attack}
\label{sec:alexa}
The contents of the user's screen can be gleaned by voice-operated virtual assistants and ''smart speakers``, such as the Amazon's Alexa assistant running on Amazon Echo devices and the Google Assistant running on Google Home devices. Once configured, such devices capture audio at all times, includign acoustic leakage from nearby screens. When a \textit{wake phrase}, such as ``Alexa'' or ``Hey Google'', is detected by the device's elaborate microphone array, it enters an attention mode to recognize and response to commands. An audio recording of the interaction, including the wake phrase utterance, is then archived in cloud servers (``to learn your voice and how you speak to improve the accuracy of the results provided to you and to improve our services''~\cite{amazonalexa}).

We set out to ascertain that the screen's visual content is indeed acoustically captured and subsequently uploaded to Google's and Amazon's cloud servers. First, we placed a Google Home device next to a Dell 2208WFPt screen displaying alternating Zebra patterns. We then woke the device up by using its wake phrase (``Hey Google'') and kept the the recording running during the Zebra alternations. Finally, we retrieved the recorded audio from Google's servers.  Figure~\ref{fig:alexa} shows the spectrogram representation of the retrieved audio, where the alternation of Zebra patterns is clearly visible. We proceeded to perform the exact same procedure with an Amazon Echo (2nd Gen.), and observed similar results.
 
\eran{SANITIZED}
\eran{SANITIZED}

\begin{figure}[t]
\iftwocol{}
\vspace{-1em}
\fi{}
\makeatletter
	\center
                \includegraphics[height=0.3\columnwidth]{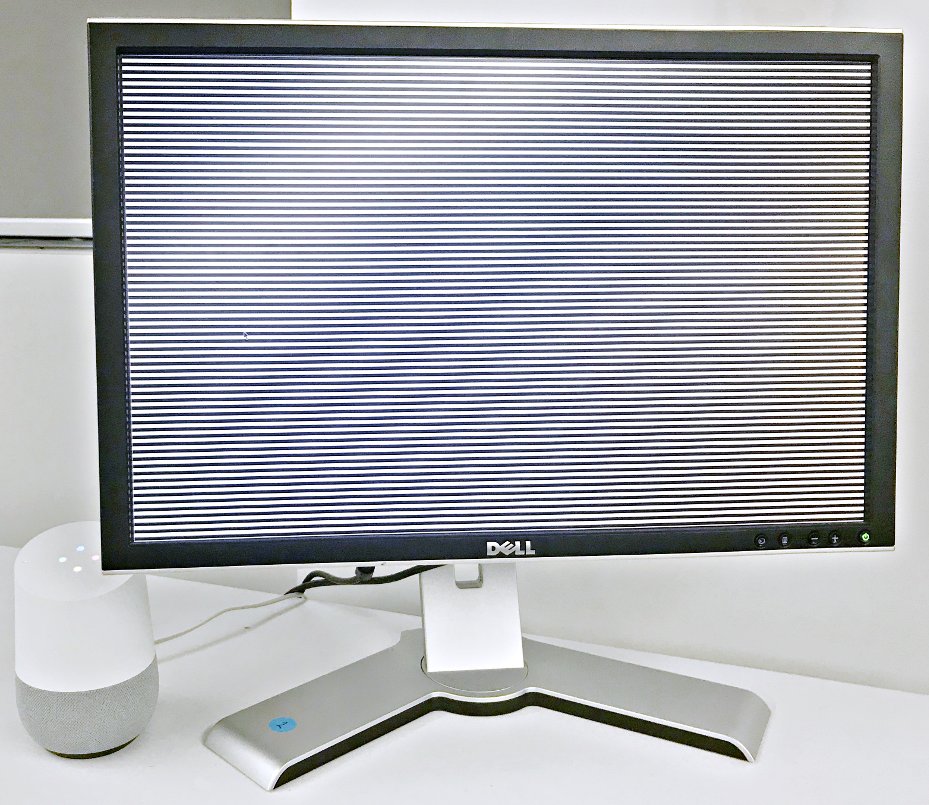}
                \includegraphics[height=0.3\columnwidth]{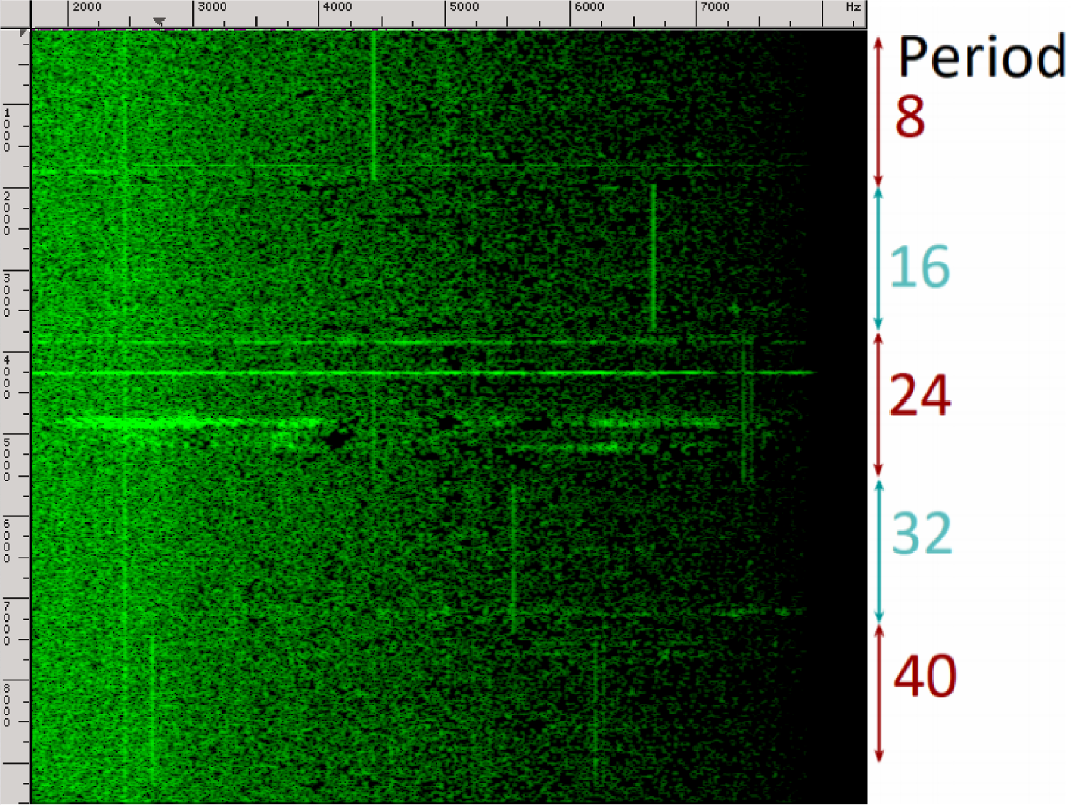}
%COMMENT
  \caption{(Left) alternating Zebra patterns displayed on a Dell 2208WFPt monitor, recorded by a Google Home device and archived on Google's cloud.\label{fig:alexa} (Right) the spectrogram of the recorded signal. Note the frequency axis ends at 8\kHz, since Google's archived recordings play at 16\kSps sample rate.}
%COMMENT
\end{figure}

\section{On-screen keyboard snooping}
\label{keyboardattack}
In this attack, the attacker aims to extract words or sentences from an on-screen keyboard. 
%COMMENT
On-screen keyboards (operated by touch or mouse) are offered by all mainstream operating systems, and also used by some websites \done\roei{SANITIZED} as a security mechanism for password entry, reducing the threat of keyloggers.%COMMENT
%COMMENT
\footnote{On-screen keyboards resist keyboard logging by malware (unless the malware is adapted to log screen content), and also low-level keyboard device snooping, e.g., by leaky USB hubs~\cite{su2017usb}.} Using an on-screen keyboard also protects against attackers with acoustic probes that try to characterize individual key acoustic emanations~\cite{asonov2004keyboard, liu2015snooping, zhuang2009keyboard}. We show, however, that the pressed on-screen keys can be inferred from the screen's acoustic emanations.

\subsection{Machine learning attack methodology}
\label{mlmethod}
Our attack works in two stages: first, in an \emph{off-line} stage, the attacker collects training data (audio traces) to characterize the acoustic emanations of a given type of screen, and uses machine-learning to train a model that distinguishes the screen content of interest (e.g., websites, text, or keystrokes). In the \emph{on-line} stage, the attacker records an audio trace of the actual screen under attack (whether in person or remotely), and then uses the trained model to deduce the on-screen content.

%COMMENT

\parhead{Why machine learning?}
The algorithm in Section~\ref{signalprocessing} produces relatively clean \textit{output traces}, where sample values are clearly dependent on the screen's content.

%COMMENT
%COMMENT
%COMMENT
%COMMENT
The correspondence between pixel intensities and sample values is intricate and difficult to accurately model\later\roei{SANITIZED}. It even seems to vary within a refresh period, with different time spans within the period presenting slightly different relationships between signal amplitude and pixel intensities.
%COMMENT
%COMMENT
Moreover, although we produce relatively clean output traces, they do contain noise. Two output traces of recordings of the same screen content will not be entirely the same. 

Neural networks, and specifically convolutional neural networks (CNNs)~\cite{cnns}, are very good at inferring such intricate dependencies in time series data, even in the presence of noise~\cite{schuster2017beauty}. In many cases, when CNNs are configured correctly, we can expect them to be able to directly perform the processing task at hand, especially for supervised learning and where enough data is at hand.

\parhead{Using classifiers.}
We employ a CNN-based architecture for solving our task. We define, train, and use CNN \textit{classifiers}: a classifier's input is a sample, represented (in our case) as a time-series vector. Specifically, our classifier inputs will be the outputs of the signal processing procedure in Section~\ref{signalprocessing}. A classifier's output is a vector of probabilities, one for each class. A sample's \textit{prediction} is the class with the highest assigned probability. Before the classifier can produce meaningful output, it has to be trained. The training procedure feeds the classifier with sample traces and correct predictions.

\parhead{Simpler ML architectures.} We fine-tuned neural networks for speed and accuracy. Alternatively, our traces (after signal processing) are often clean enough to train a rudimentary logistic regression model (at the cost of much slower training, due to slow convergence). For the lower-frequency non-modulated signals acquired through VoIP and used without trace averaging (Section~\ref{codecattack}), the rudimentary model attains low accuracy, but a carefully tuned CNN is effective.

\subsection{Attack simulation and results}
\label{sec:keyboardsnoop1}

We simulated this attack in the smartphone and close-range attack vectors. We assume the on-screen keyboard is sufficiently large, depending on screen attributes\later\roei{SANITIZED}.

\parhead{Data collection and preprocessing.}
We captured audio recordings of a Soyo 20'' DYLM2086 in the close-range and smart phone settings (Section~\ref{sec:closevsfar} and~\ref{sec:phoneattack}, while displaying screen shots of the mouse pressing varying keys. We used the native Ubuntu on-screen keyboard with a phone-style layout, ``High Contrast'' theme and size 700x900. We used portrait screen layout (see discussion below on screen layouts).

For each attack vector, we iterated 100 times over the 26 letters of the alphabet. To reliably simulate the acoustic leakage during hundreds of key-presses on the on-screen keyboard required for training, we used screenshots containing the on-screen keyboard with the respective key being pressed. In each iteration, for each key, we collected a 0.5\second long trace when key is pressed (simulated by the screen shot). We split these traces to train (90\%) and validation (10\%) sets.~\footnote{Here and in Section~\ref{textattack}, we use just 10\% of the initial recordings for validation. In both, validation sets are still in the hundreds, and we record an additional trace set, 100\% of which was used for testing.}

For testing our snooping attack, we also recorded audio while words are typed on the virtual keyboard. Words were chosen randomly from from a list of the most common 1000 English words~\cite{commonwords} (see results). We use the traces of 50 such words for testing both classifiers. Characters of the words were typed consecutively, with 3\second between each transition. We also experiment with words recorded at 1\second speed (see results).

%COMMENT

We applied our signal processing procedure in Section~\ref{signalprocessing}.

\parhead{Training procedure}
Trace processing and training was performed on an Intel(R) Xeon(R) CPU E5-2609 v4 processor with two Titan X GPUs. We uses the Keras Python library with a Tensorflow backend to program and train the networks. Keras is used with a Tensorflow 1.4 backend and CUDA 8.0, CUDNN 6.0. We defer the details of the neural network's architecture and hyperparameters to Appendix~\ref{sec:cnnsnoop}.

Each on-keyboard keystroke results in a specific image displayed on the screen. We can train a classifier to identify the different characters for a given trace. Some pairs of keys are completely horizontally aligned on this virtual keyboard. We expect keys in such pairs to be less easily distinguishable from the signal (see below discussion about screen layouts), and group each such pair into one class label. Our labels were thus `b', `c', `m', `n', `p', `v', `x', `z', `aq', `sw', `de', `fr', `gt', `hy', `ju', `ki', `lo', and space.

\parhead{Testing procedure.}
To process a trace, we shift a 0.5\second window across the duration of the trace. The window offset is advanced in 3200 sample ($1/60\second$) intervals. For each offset, we apply our processing algorithm on the sub-trace contained in the window. We then apply our classifier.

This process outputs an array of class labels of size $60\cdot l - 30$, where $l$ is the original trace length. We traverse it from start to finish, outputting class labels appearing more than 15 times contiguously for the smart phone attack and 35 times contiguously for the close-range attack\later\roei{SANITIZED}. Assuming the CNN always predicts the right class, this produces sequences of class labels that correspond to the letters typed on the screen. However, these sequences do not distinguish (1) letters grouped into the same class, and (2) letters in the same class that were typed sequentially. For example, for a trace of the word ``love'', the sequence ['lo', `v', `de'] will be produced. The `l' and `o' keys are grouped into the same class, and there is only one instance of this class label. Similarly, the expected trace of the word ``screen'' is ['sw', `c', `fr', `de', `n'].

To disambiguate each produced trace, we go over the 102000 words in the dictionary. For each, we compute its expected trace, and check it against the trace. We return the list of words that matched the trace, or the \textit{prediction list}.

\parhead{Results.}
Our classifier reaches 100\% accuracy on the validation set, for both the smartphone and close-range.  For the close range attack's word traces, the correct word was always contained in our prediction list. The size of the prediction list varied from 2 to 23 with an average of 9.8 words per trace. For the smartphone attack, the correct word was contained in the prediction list for 49 out of 50 traces.

To further test the limits of this attack, we repeated the attack using a different screen, Soyo 22" (model DYLM2248), in the close-range setting. Again, the validation set accuracy was 100\%. This time, we collected traces of 100 additional words, twice: once while waiting 3 seconds between each transition, and once when waiting 1 second. For the 3\second collection, the correct word was in the prediction list 94 times out of 100. For the 1\second recordings, the correct word was in the prediction list 90 times. The average candidate list size was 8.

\parhead{Portrait vs. landscape layout.}\label{portraitvslandscape} A pressed key's color changes from white to black; we expect this to affect the amplitude of the signal emitted while the corresponding black pixels are rendered. Pixels are rendered in a raster pattern, i.e., line by line. Because of that, keys that are horizontally aligned, and span the same pixel lines, are rendered during very similar times in every refresh period. We therefore expect that pressing horizontally-aligned keys will induce very similar, temporally-close effects in the leakage signal. Conversely, keys positioned in completely different pixel-lines will be rendered during very different times, and the corresponding amplitude changes will be more distinguishable.

For this reason, in this section, we first built a classifier that only distinguishes between non-aligned keys, and test it on a portrait layout where far fewer keys are horizontally aligned.

%COMMENT

%COMMENT

To ascertain whether horizontally-aligned keys can be at all distinguished, we repeated data collection using the Soyo 22" in landscape layout. We trained the classifier on both landscape and portrait datasets, but without grouping horizontally aligned keys. The portrait layout classifier reaches 96.4\%, and the landscape layout classifier reaches 40.8\%. Top-3 accuracies are much higher: 99.6\% and 71.9\%. We conclude that even changes \emph{within} pixel lines, where pixels are rendered at very small temporal offsets from each other, can still be distinguishable in the signal. Moreover, that on-screen keyboards remain vulnerable even in landscape layouts.

\parhead{Cross-screen results.}
We verified that the attacker's training can be done on a different screen than the victim's. To this end, we used a smartphone to  collect about 130 recordings, 0.5\second each, for each key on a Dell 2208WFPt screen in portrait layout. We then switched to a different 2208WFPt instance, and collected 10 traces, 0.5\second each, per key. We followed the same preprocessing procedure as above, and assigned classes to traces, again grouping horizontally-aligned keys. We then trained our classifier using the traces from the first screen and tested it on the traces of the second screen. The resulting accuracy is 99.0\%.
For comparison, this classifier had a 99.4\% accuracy on a similar-sized validation set of traces of the first screen (not used for training). We conclude that the keyboard snooping attack in the smartphone setting is possible for a remote attacker without access to the victim's screen. This supports the results of Section~\ref{sec:crossscreenmany} which indicate that using more than one screen for training would likely result in even better generalization.

\later\roei{SANITIZED}

%COMMENT
%COMMENT
%COMMENT
%COMMENT
\section{Text extraction}
\label{textattack}
In this attack, the attacker aims to extract text from the attacked screen. This could be used to extract various types of sensitive data, such as identifying information, private messages, search queries, and financial information.

 In this attack, we explore the possibility of extracting content from the screen rather than classifying it. We simulate an \textbf{open-world setting} where the word base rate is as low as $1/55000$, i.e., a specific word has a $1/55000$ probability of appearing in a trace. To simulate the low base rate, our implementation assumes that all characters and words are equally likely: it does not use any knowledge of actual word or character distribution. Notably, this is conservative for evaluating natural language word extraction, where the average base rate is much higher in practice. 

We simulate an attack on a portrait-layed victim screen displaying black-on-white text in very large monospace font (we discuss these assumptions in Section~\ref{sec:limitations}).

\parhead{Data collection and preprocessing.}
We layed a Soyo 22" DYLM2248 in portrait layout, and captured traces in the close-range setting (see Section~\ref{sec:closevsfar}). We collected 10,000 traces, each 5\second long. In each trace, a different sequence of randomly chosen letters was displayed on the screen. The length of sequences was chosen randomly from $\{3,4,5,6\}$. Letters were chosen from the English alphabet. Letters were capitalized, in font type Fixedsys Excelsior and size 175 pixels in width. Letters were black on a white screen. We attach each trace to its corresponding sequence of characters, and again split the traces to train (90\%) and validation (10\%), and employ the same machine learning methodology as in Section~\ref{mlmethod}. 

Similarly, we also collected a test set: the traces of 100 English words, 25 for each possible length, chosen randomly similarly to Section~\ref{sec:keyboardsnoop1}. We apply the signal processing algorithm described in Section~\ref{signalprocessing}.

\parhead{Training procedure.}
We build one classifier for each character location. Each character location is rendered at a specific time segment during each refresh cycle. We match every character location with its time segment.%COMMENT
%COMMENT
\footnote{To measure this, we introduced sharp pixel intensity changes in two different pixel lines and measured when, during the refresh cycle, these changes affect the signal. Because line rendering time is linear in line numbers, we can use this to construct an accurate mapping of line\full{SANITIZED}s to rendering time.} We construct training and validation data for each character by extracting the respective output trace segment, thus collecting pairs of output trace segment and corresponding character value.

Appendix~\ref{sec:cnntext} details the CNN's architecture. We again trained it on a GPU-enhanced machine (see Section~\ref{sec:keyboardsnoop1}).

\parhead{Testing procedure.}
For each of our 100 test set words, each classifier outputs a probability vector for each character. To predict which word was typed in, we use the Webster's dictionary~\cite{gutenberg}. We extract the first 6 characters of each dictionary word, sum the log-probabilities of each word's characters, and output the words sorted by their probabilities.
%COMMENT
%COMMENT

\parhead{Results.}
The per-character validation set accuracy (containing 10\% of our 10,000 trace collection) ranges from 88\% to 98\%, except for the last character where the accuracy was 75\%. Out of 100 recordings of test words, for two of them preprocessing returned an error. For 56 of them, the most probable word on the list was the correct one. For 72 of them, the correct word appeared in the list of top-five most probable words.

\parhead{Error analysis.}
Our attack often confuses a commonly used word for a rare one. For example, the word ``dream'' is erroneously predicted as ``bream''. To avoid such errors, we can introduce priors on word and character distributions (increasing our conservatively low base rate). For example, by assigning a higher probability for frequent words in the probability-assigning phase. This would likely significantly increase accuracy, especially for commonly used words such as those in our training set. 

\section{Website distinguishing}
\label{websitedistinguish}
In this attack, the attacker is interested in learning whether the victim is entering a specific website, or is just interested in the victim's website visiting habits.
\full{SANITIZED}
Website fingerprinting attacks, often studied in network traffic analysis settings~\cite{panchenko2016website, schuster2017beauty, cai2012touching}, convey the target user's browsing behavior and are very revealing of their hobbies and personality~\cite{kosinski2014manifestations}. In the VoIP setting, the attacker may be interested to learn in real time what the other parties to the call are doing. For example, he may wish to learn whether another party to the call is currently keeping the videoconference app maximized (presumably focusing his attention on the call), browsing Facebook, or responding to E-mails.

We note here that the classifiers described in Section~\ref{modulatedwebsite} accurately distinguish up to 100 websites. Thus, as our methodology can accurately classify screen content, it can be leveraged for detecting an identified website visit vs. using common apps as well as classifying common apps. Finally, we also note that the collected website traces naturally contain dynamically-changing content (e.g., ads) affecting measurements, but classifiers nevertheles attain the high accuracy (implying that the general layout of a website is typically static and unique).

In Section~\ref{codecattack}, we directly evaluate a VoIP attacker whose goal is identifying the victim's foreground window, distinguishing between a VoIP app and popular websites.

 \later\eran{SANITIZED}\later\roei{SANITIZED}
\full{SANITIZED}

\subsection{Using the modulated signal}
\label{modulatedwebsite}

\parhead{Data collection and preprocessing.}
We recorded traces of a Soyo 22" DYLM2248 at the close-range, at-distance, and phone attack settings (Sections~\ref{sec:closevsfar} and~\ref{sec:phoneattack}). Out of the Moz top-500 list~\cite{moz}, that ranks websites according to the number of other websites linking to them (which highly correlates with popularity), we chose 97 websites by filtering out corner cases.  For example, duplicate domains for the same website (e.g., \url{google.co.in} and \url{google.co.jp}) and non-US-based websites because some of them cause Selenium to hang. 
%COMMENT

We simulated the attack for the smartphone, at-distance, and close-range vantage points. For each, we iterated over the collection of websites over 100 times. In each iteration, we opened a Chrome browser window (controlled by the Selenium automation library for Python) with the URL of the website. We then started recording for 5\second. For each vantage point, we collected traces for 5 consecutive nights (when the recording machine was not otherwise in use). We stopped when we had reached 100 samples per website. This resulted in about 130 traces per website in the close-range vantage point, 100 traces from the at-distance vantage point, and 110 traces in the smartphone vantage point. For each setting, we used 70\% of traces for training. For the close-range setting, we used the remaining 30\% as a validation, which we used to guide our classifier architecture tuning (e.g., set learning rate, number of convolutional layers, etc). For the at-distance and smartphone settings, we used the remaining 30\% as test sets. 
\ignore[Beyond 70\%]{
  We found that in the close-range setting, increasing the train set size beyond 70\% does not yield significant improvement in results.
}
We apply the signal processing algorithm described in Section~\ref{signalprocessing}.

\parhead{Machine learning and results.}
Our task is to find which of 97 websites is displayed. We train a CNN directly to solve this task (see Appendix~\ref{sec:cnnwebsite}), using the setup from Section~\ref{sec:keyboardsnoop1}. In about 8\% of traces in the close-range and phone attacks, and about 16\% in the at-distance attack, the signal processing algorithm returned an error, implying the signal is particularly noisy. For the close-range setting, the validation set accuracy was 97.09\%. For the smartphone and at-distance test sets, the accuracy was 91.20\% and 90.9\% respectively.\later\roei{SANITIZED}\later\roei{SANITIZED}

\parhead{Eliminating false positives.} \done\daniel{SANITIZED}\done\roei{SANITIZED}
Our classifiers in this section work in a closed-world, where the victim visits one out of the (here, 97) targeted websites. In reality the victim may also visit other, unknown websites, and the attacker may be interested in detecting when that occurs.
The victim could visit numerous websites before ever visiting a targeted website, so even an ostensibly low probability of an alarm on a single non-targeted site's trace, can lead to many false alarms. This is so-called ``base rate fallacy'' is extensively discussed in fingerprinting literature \cite{schuster2017beauty, juarez2014critical}. We thus aim to further minimize the probability of false identification (false positive rate).

%COMMENT

\done\roei{SANITIZED} Neural network classifiers assign to all classes values between 0 to 1, whose sum is 1 (similar to a probability distribution). The prediction is the class with the highest value, which can be interpreted as the prediction's confidence. We can prioritize precision over recall by dropping predictions where confidence is below a threshold. By setting the threshold to 0.96, we get precision of 0.996, whereas the recall remains well above 0.94. Our classifier is confidently erroneous only on 14 out of 3222 validation set samples. Moreover, for every other sample where a confident mistake was not made, it successfuly disqualifies not 1 class, as the attacker above requires, but 96 classes. In other words, it makes only 14 ``confident'' mistakes out of $3222\times 96 = 309312$ possible ones (it is confidently mistaken at a rate of $\approx 5/100000$). Thus, while we do not directly simulate an open-world setting, our results do demonstrate the necessary low amount of false positives for detection in very low base rates.

\parhead{Cross-screen results.}
In Section~\ref{sec:crossscreenmany} we extensively evaluate the prospects of this attack in a cross-screen setting, where the attacker has no access to the victim's screen.

\subsection{Attack through a Hangouts call}
\label{codecattack}
Here, we assume the attacker and victim are sharing a Hangouts video-chat session, where audio from the victim's environment is transmitted directly to the attacker. Leveraging acoustic leakage from the victim screen, the attacker's goal is to distinguish the scenario where video-chat is at the foreground from a scenario where the victim is surfing the Web, and also tell what website they are visiting.

\parhead{Data collection and preprocessing.}
We recorded traces using the setup described in Section~\ref{sec:voip}.
We iteratively switched the foreground window of the victim screen in a round-robin fashion between 11 open windows: browser windows for each of the 10 top websites in Moz, and a (screen-shot) of a video-chat showing the human face of the attacker sitting in front of his webcam. We captured a 6\second recording before switching the next window into the foreground. In this way we collected 300 traces, 6\second each, for each open window.

In previous attack settings, we used the fact that the pattern of interest is modulated and transmitted, in every refresh period, over a carrier signal at 32\kHz. We leveraged this to produce a relatively clean version of a display-content-dependent leakage signal (see Section~\ref{signalprocessing}). Here, we only sample at 44\kSps ---below the Nyquist rate of the carrier signal. We can still, however, leverage the effect described in Section~\ref{sec:timedomainanalysis}---namely, that pixel intensity directly affects the acoustic signal's amplitude. We process traces in a more straightforward way: we computed the fast Fourier transform of each 6\second trace. This results in a vector of frequency coefficients, corresponding with frequencies between 0\kHz and $44100/2=22050\kHz$. We take coefficients of frequencies in band 9-15\kHz which, we found, contain sufficient information, and used them as the classifier input vector. We split the traces to train (70\%) and validation (30\%).
%COMMENT

\parhead{Machine learning and results.}
Our task is to find which of 11 \done\eran{SANITIZED} windows was in foreground. As in Section~\ref{modulatedwebsite}, we design a CNN directly trained to solve this task (see Appendix~\ref{sec:cnnwebsite}). The CNN reaches 99.4\% accuracy on the validation set.

We compare results on over-VoIP acquired traces and traces acquired in the close-range setting. To facilitate a fair comparison, we need 2 similar datasets of traces, with the difference that one was collected through the close-range setting and preprocessed as in Section~\ref{modulatedwebsite}, and one colleted via the VoIP setting and preprocessed as above. We use subsets of the datasets described in this section and that in Section~\ref{modulatedwebsite}; both subsets contain about 80 5\second training traces per website. We trained a classifier for each dataset, using the appropriate CNN architecture (see appendix~\ref{sec:cnnwebsite}). The through-VoIP classifier attained an accuracy of 98\% on the validation set, slightly less than the through-VoIP classifier described above (which handles 1 extra class, but uses x3 more traces per class). The close-range classifier attained an accuracy of 99.2\%.

We conclude there is a minor drop in the attacker's accuracy when sound is recorded using commodity equipment and is moreover encoded and decoded using a lossy VoIP codec. However, using a higher amount of traces, we still attain near-perfect accuracy in the through-VoIP setting.\later\roei{SANITIZED}

\parhead{Cross-screen results.}
We collected and processed 30 traces per class (totaling 330) from another Dell 2208WFPt screen, and tested our classifier on them. Our classifier had an accuracy of around 0.1, slightly above a random guess, indicating that the classifier is overfitted to the training screen. In Section~\ref{sec:crossscreenmany}, we show that overfitting can often be mitigated by training on more than one screen. We leave the task of applying that methodology for this attack to future work.

\section{Cross-screen attacks}
\label{sec:crossscreenall}
\label{sec:crossscreenmany}
Thus far, we performed preliminary evaluations of the cross-screen scenario for most attacks, by training on one screen and testing on a different one. While this proved highly effective in some cases, such an attacker does risk training a model oferfitted to the traits of a specific screen. In this section we demonstrate how overfitting can be mitigated by training on multiple screens rather than one.

\parhead{Data collection.}
\newcommand{\Set}[1]{\textsf{\textit{#1}}}
\newcommand{\Screen}[2]{\textsf{\textit{#1\##2}}}
We used a total of ten screens: five of model Dell 2208WFPt WD-04 (the \Set{Dell4} set), two of model Dell 2208WFPt WD-05 (\Set{Dell5}), two of model Dell 2208WFPf B (\Set{DellB}), and one of model Soyo 22" DYLM2248. The screen set was chosen to contain both similarity (nine screens in the Dell 2208WFP* family) but also variation (including three different Dell models and one Soyo model).  For every screen, we collected 50 recordings, 5\second each, of each of 25 websites (similarly to Section~\ref{websitedistinguish}, the top 25 websites in the Moz top 500), in the close-range setting (see Section~\ref{sec:closevsfar}). 

\parhead{Data preprocessing.}
We first applied the signal processing from Section~\ref{signalprocessing}. Then, for each victim screen $v$, we evaluate classifiers trained using each of its \textit{training collections}:
\iftwocol{}
\begin{itemize}[nosep,leftmargin=*]
\else{}
\begin{itemize}
\fi{}
  \item Each single screen (including $v$).
  \item Each of same-model sets (\Set{Dell4}, \Set{Dell5} and \Set{DellB}) defined above, excluding $v$ from the corresponding set.
  \item The \Set{mixed} collection, containing 2 randomly chosen screens from \Set{Dell4}, 1 from \Set{Dell5}, and 1 from \Set{DellB}, excluding $v$.
  \item The \Set{all} collection, containing all 10 screens, excluding $v$.
  \item The \Set{nosoyo} collection, with all screens except the Soyo.
\end{itemize}

\noindent For every such collection $c$, we assembled a training data containing about 50 samples for each website, by taking $50/\left|c\right|$ samples for each screen in $c$. 

\parhead{Machine learning and results.}
We used our website distinguisher architecture similar to that in Section~\ref{websitedistinguish}, but used the Adadelta optimizer which converges much faster than SGD when using only 25 classes, and thus trained for only 200 epochs. For each $v$, we trained our classifier on each of its training collections and evaluated the resulting classifier on $v$'s data (re-initializing learned weights at random before the next training process). The results are summarized in Figure~\ref{fig:heatmap}. 

\begin{figure}[t]
\makeatletter
\vspace{-2em}
\centering  \includegraphics[width=0.8\columnwidth]{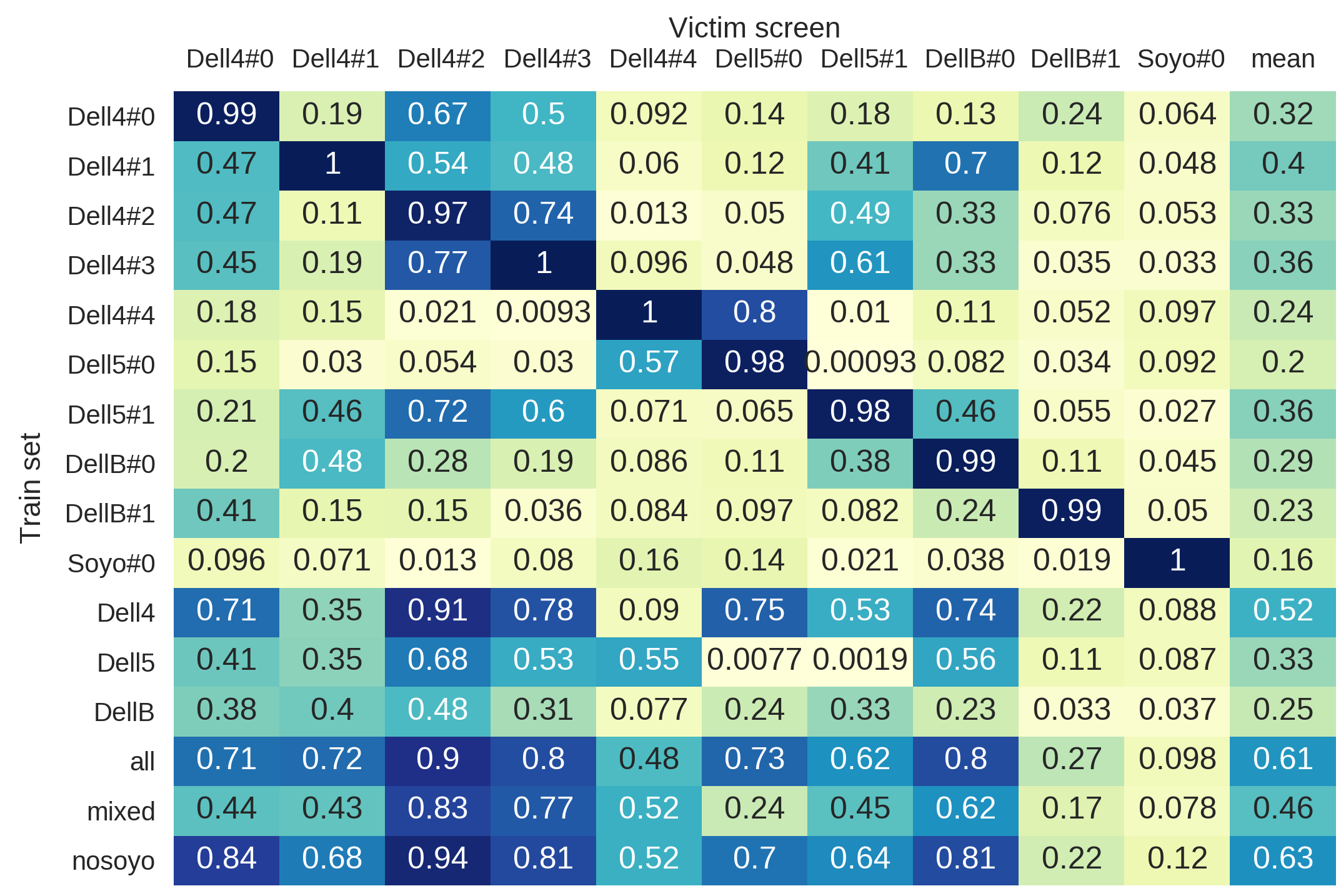}
%COMMENT
  \caption{Cross-screen classification accuracy.\label{fig:heatmap}}
  \vspace{-1em}
\end{figure}

\parhead{Observations.}
First, we observe that classifiers trained on one of the 3 Dell models often generalized to a different model within the Dell 2208WFP* family. This sometimes happens even when training on just one screen (e.g., several classifiers generalize well on screen \Screen{Dell4}{2}), and especially when training and testing on screens of the same model. Second, using more screens from the same family improves the intra-family generalization:  training on a single screen yields worse results than training on two screens (i.e., \Set{Dell4} or \Set{Dell5}); training on 4 screens yields further improvement (\Set{mixed} and \Set{Dell4}); and training on 9-10 screens (\Set{all} and \Set{nosoyo}) gives the best results. Third, intra-model generalization is slightly higher than generalization across the Dell models: for classifiers trained on a single Dell screen and tested a different screen of the same model, the average accuracy is 0.276, compared to 0.233 for screens of other Dell models. Finally, inter-vendor generalization is poor: classifiers trained on the Soyo screen have low accuracy for Dell screens and vice versa.

\parhead{Conclusions.}
The accuracy of a remote attacker with no physical access to the screen is limited by inter-screen generalization. To attain high generalization, the attacker can use multiple screens from the same model, or even similar models from the same vendor. 
%COMMENT
Note that this training phase can be done once, off-line, and utilized for multiple attacks. It can also be done retroactively, after recording the victim and using this recording to fingerprint their screen model.
\later\eran{SANITIZED}

\ignore[Cross-vendor]{
Furthermore, since some leak characteristics are consistent across models, families, and vendors, it is likely that introducing yet more variation into the training set will yield better inter-vendor results. This would allow an attacker to train once and attack multiple victim models, or facilitate even an attack using no knowledge of the victim screen's model, family, or vendor.
}

%COMMENT

 \later\eran{SANITIZED}
%COMMENT

\begin{comment}
\parhead{On-screen keyboard snooping.}
On-screen keyboards are considered safer against adversaries with only audio capabilities. We show that, in fact, the opposite is the case. In fact, such an adversary can extract entire words with an extremely high recall and tractable precision.

\parhead{Open-world extraction.}
Close-range website distinguishing works with extremely high accuracy (0.97) for distinguishing 97 popular websites. Comparatively, our text extraction attack performs a task harder than classification: it \emph{extracts} screen content that it was not directly trained to identify. For example, our training set does not contain the word ``build'', but it is nevertheless correctly identified. This stems in our ability to predict characters individually, using just portions of the traces that correspond to their rendering times.

The feasibility of extracting content with extremely low base rate ($1/55000$) and with high confidence (>0.5) indicates that enough is leaked to expose sensitive user information. Values filled by users in HTML forms, such as search queries and credit card details, are particularly exposed to such an attack since the screen layout could be known to the attacker.
\end{comment}

\section{Limitations}
\label{sec:limitations}
While the results presented in this paper clearly demonstrate that screen content can be detected via an acoustic side channel, the presented attacks have some limitations:

\parhead{Remote attack accuracy.}
Accuracy is reduced in a remote setting where the adversary uses their own screen(s) for training (see Section~\ref{sec:crossscreenall}, Section~\ref{sec:keyboardsnoop1}, and Section~\ref{modulatedwebsite}). However, in many cases it remains high (e.g., over 90\% for distinguishing between 25 websites, 98\% for distinguishing letters typed on an on-screen keyboard). Our investigation in Sections~\ref{sec:crossscreenzebra} and~\ref{exploringleakage} indicates that different screens display similar signal content-dependence, explaining why attacks generalize across screens (when enough screens are used for training).

\parhead{Attacker's prior knowledge.}
The adversary needs to know the victim's screen model (or face reduced accuracy). Note that such knowledge is far less sensitive than the screen's actual content, and is often readily deducible from the victim's purchasing procedures or trash. Changes in components other than the screen could also affect the acoustic leakage (e.g., the attached computer may use the screen at an unusual display resolution), but we conjecture that in practice this is rare.

\later\roei{SANITIZED}

Moreover, for on-screen keyboard snooping (Section~\ref{sec:keyboardsnoop1}), an attacker needs to know the on-screen layout of the victim, including text background, font, key size, etc.  We consider this scenario plausible, as website and keyboard layouts as well as fonts are often fixed (e.g., when the victim uses the OS's native on-screen keyboard to type a login password). Similarly, our text extraction attack currently requires a known background and a large 175pt font. Finally, we note that our website classification experiments assumes the simple case of  a user statically viewing a webpage. We leave the task of achieving website classification in the presence of user interaction (such as active scrolling) as future work. 

\parhead{Granularity.}
Our attacks detected information at a a coarser resolution than individual pixels. Indeed, the requisite bandwidth for detecting individual pixels seems to far exceed the acoustic transmission properties of air (modern screens render pixels at a rate of over 100\MHz, whereas above a few hundred kHz sound propagation in the air has too short a range due to attenuation and distortion effects).

This affects the type of screen contents that can be distinguished. For example, our text extraction (Section~\ref{textattack}) works on large fonts. Similarly, we show that on-screen keyboard snooping (Section~\ref{portraitvslandscape}) is more accurate in portrait orientation, where coarse pixel \emph{line} granularity suffices.\later\eran{SANITIZED}

\parhead{Attacker sample rate.}
In Sections~\ref{sec:keyboardsnoop1} through~\ref{modulatedwebsite}, we exploit a carrier signal at 32\kHz. To carry out these attacks, the attacker's sample rate must be at least 64\kSps (due to the Nyquist limit, and assuming that aliasing is eliminated by hardware filters). While modern commodity hardware often samples at 96\kSps (as we demonstrated in Sections~\ref{sec:phoneattack},~\ref{sec:keyboardsnoop1},~\ref{modulatedwebsite} using a mobile phone), low-end attackers recording at lower rates might be limited to the remote attacks demonstrated in Section~\ref{codecattack}.

\parhead{Accuracy depends on microphone distance and screen model.}
The leakage signal quality and extractability varies greatly with the microphones' proximity to the screen (see Figure~\ref{phones}), as well as screen makes and models (see Figure~\ref{fig:monitor-many}). However, as we show, even relatively noisy signals acquired at-distance (Section~\ref{modulatedwebsite}), or signals passed through a lossy codec (Section~\ref{codecattack}), can be used to mount an attack. Deterioration of the underlying signal with distance is analyzed in Appendix~\ref{sec:distancevsquality}.

\later\roei{SANITIZED}

\section{Mitigations}
\label{sec:mitigations}
\parhead{Elimination}
An obvious remedy to such leakage is for computer screen manufacturers to more carefully evaluate their designs, to minimize ``coil whine'' and similar electronic component vibrations within screen circuitry (cf.~\cite{GST14acoustic-norefs}). The ubiquity of leakage, across manufacturers and models (Figures~\ref{fig:monitor-many},~\ref{phones},~\ref{fig:webCams} demonstrate leakage in Dell, Soyo, Apple, Philips, HQ, BenQ, and Samsung; attacks were simulated on various Dell and Soyo screens in Sections~\ref{keyboardattack} through~\ref{sec:crossscreenall}), suggests that this may be difficult or costly. 

\parhead{Masking}
Acoustic noise generators can be used to mask the signal, at a cost in design, manufacturing, and ergonomic disruption (some of the exploitable signal lies within the human-audible frequency range).
The masking ought to have adequate energy and spectrum coverage to reduce signal-to-noise by orders of magnitude, because the leakage signal is retransmitted 60 times per second, offering high redundancy. 

\parhead{Shielding}
Acoustic shielding of screens may reduce the leakage amplitude, but is difficult to achieve while keeping adequate air circulation for cooling. For microphones built into screens, a sound-absorbing barrier may reduce microphone pickup of internally-generated sounds (but would not affect external microphones). 
In the case of some screens or microphones in which there is an electromagnetic contribution to the leakage (see Section~\ref{sec:acoustic-or-em}), corresponding shielding would also be desirable---and, typically, expensive.
%COMMENT
%COMMENT
%COMMENT

%COMMENT
%COMMENT

\parhead{Software mitigations.}
A more promising approach to mitigating the attacks presented in this paper are software countermeasures. More specifically, variations on software mitigations to the EM Tempest attack, which change on-screen content to mask the leakage, such as font filtering~\cite{kuhn1998soft}, can be considered. Moreover, because our extraction attacks use mainly aggregate horizontal intensity of pixel lines, while mostly losing the information inside individual lines, fonts can be changed such that all letters project the same horizontal intensity. Finally, our attacks all heavily rely on neural network classifiers which, themselves are vulnerable to inputs specifically crafted to mislead them~\cite{goodfellow2014explaining, szegedy2013intriguing}. 

\section{Conclusion}
\label{conclusions}
We report a new physical side-channel leak: acoustic leakage of screen content. We suggest \emph{powerful attacks} that extract highly precise information, such as on-screen text and on-screen keyboard presses. We posit that this leakage is uniquely dangerous because even \emph{weak attackers}, that only receive encoded audio traces from legitimate communication channels, as well as a attackers with no access to the victim's physical screen, can exploit it. 

We demonstrate this by successfully simulating highly precise content extraction and identification attacks across an array of setups, as well as a simple, but well motivated, attack scenario: exploiting an open Hangouts connection to deduce what on-screen activity a party to the call is involved in.

This is the first demonstrated attack using codec-encoded acoustic emanations from non-mechanical peripherals, for which users don't have a reason to suspect acoustic leakage would even exist. 

\iftwocol{}
\vspace{-1em}
\fi{}

%COMMENT
\section*{Acknowledgments}
\iftwocol{}
Roei Schuster and Eran Tromer are members of the Check Point Institute for Information Security. 
\else{}
\fi{}
This work was supported
%COMMENT
by the Blavatnik Interdisciplinary Cyber Research Center (ICRC); %COMMENT
by the Check Point Institute for Information Security; %COMMENT
by the Defense Advanced Research Project Agency (DARPA) and Army Research Office (ARO) under Contract \#W911NF-15-C-0236; %COMMENT
%COMMENT
by the Israeli Ministry of Science and Technology; %COMMENT
by the Leona M. \& Harry B. Helmsley Charitable Trust; %COMMENT
by NSF awards \#CNS-1445424 and \#CCF-1423306; %COMMENT
by the 2017-2018 Rothschild Postdoctoral Fellowship; %COMMENT
and a gift from Intel.
%COMMENT
%COMMENT
Any opinions, findings, and conclusions or recommendations expressed are those of the authors and do not necessarily reflect the views of ARO, DARPA, NSF, the U.S. Government or other sponsors.
%COMMENT

\iftwocol
  %COMMENT
  \done\roei{SANITIZED}
  \bibliographystyle{IEEEtranS}
\else
  \bibliographystyle{alphaurl}
\fi

%COMMENT

\bibliography{references}

\appendix

\section{Appendix}
\subsection{Our trace chunking algorithm}
\subsubsection{Algorithm in detail}
\label{chunkingalgo}
The algorithm first initiates an empty collection of chunks. Then, it searches for the first two consecutive chunks whose sizes are in $G$, and their correlation\footnote{Pearson correlation of two chunks of different sizes is taken after truncating them to the lower size among the two.} is higher than the threshold. It sets one of these chunks to be the first, master chunk, and adds it to the collection.

Then, starting from the position after the added chunk, it searches a $size\in G$ for the next chunk such that the correlation of the \emph{following} chunk is the highest. If the master's correlation with the obtained chunk is above $T$, it is added to the chunk collection. Otherwise, it is discarded and the algorithm goes into \textit{sync mode}.

In sync mode, the next chunk's size can be between $S$ and $S+6000$ samples. Again, the algorithm finds the size that maximizes correlation with the following chunk. As long as the algorithm is in sync mode, it does not add new chunks to the collection. The algorithm exits sync mode once it found a size such that the following chunk has correlation $<T$.

When the algorithm succeeds, it truncates all chunks to the size of the smallest one (typically $min\{G\}$), and performs outlier rejection: discard the 10\% of chunks whose correlation with the master chunk is lowest, as well as any chunk that has a peak that exceeds 1.5 times the average highest peak.
\full{SANITIZED}
The pseudo-code is given in Algorithm~\ref{algochunk}.

%COMMENT

%COMMENT
%COMMENT
%COMMENT
%COMMENT
%COMMENT
%COMMENT
%COMMENT
%COMMENT
%COMMENT
%COMMENT
%COMMENT
%COMMENT
%COMMENT
%COMMENT
%COMMENT
%COMMENT
%COMMENT
%COMMENT
%COMMENT
%COMMENT
%COMMENT

%COMMENT
%COMMENT
\newcommand{\algstep}[1]{\textrm{#1}}
\newcommand{\identifier}[1]{\textsf{#1}}
\newcommand{\makeIdentifier}[1]{%COMMENT
  \expandafter\newcommand\csname#1\endcsname{\identifier{#1}}}
\makeIdentifier{chunks}
\makeIdentifier{len}
\makeIdentifier{state}
\makeIdentifier{corr}
\makeIdentifier{signal}
\makeIdentifier{argmax}
\makeIdentifier{median}
\makeIdentifier{append}
\newcommand{\IDlist}{\identifier{list}}
\newcommand{\clen}{\identifier{c\_len}}
\newcommand{\nextlen}{\identifier{next\_len}}  %COMMENT
\newcommand{\synccount}{\identifier{sync\_count}} 
\newcommand{\outlierreject}{\identifier{outlier\_reject}}
\newcommand{\normal}{\texttt{"normal"}}
\newcommand{\STsync}{\texttt{"sync"}}
\newcommand{\STnormal}{\texttt{"normal"}}
\newcommand{\STerror}{\texttt{"error"}}

\begin{center}
%COMMENT
	%COMMENT
	\captionof{algorithm}{\algstep{Chunkify}\label{algochunk}}
	\scriptsize
	\begin{algorithmic}[1]
		\Procedure{ChopSignal}{array $\signal$, size $S$, allowed drift $d$, threshold $T$}
		\BState \algstep{init}:
		\State $\textit{\chunks} \gets \IDlist()$
		\State $G \gets \left[S-d-1, S+d\right]$
		\State $\clen \gets \median\{G\}$ // expected chunk length
		\\

		\BState \algstep{find\_first}, \algstep{master\_chunk}:
		\While {$\len(\chunks)=0$}
			\State $j \gets \argmax_{j\in G}{\{\corr(\signal[:j], \signal[j:j+\clen])\}}$
			\If {$\corr(\signal[:j], \signal[j+1:])>T$}
				\State $\chunks[0] \gets \signal[:j]$ // master chunk
				\State $\signal \gets \signal[j:]$
			\Else
				\State $\signal \gets \signal[1:]$
			\EndIf
		\EndWhile

		\BState \algstep{loop}:
		%COMMENT
		\State $\nextlen \gets G$ //normal mode
		\State $\state \gets \STnormal$
		\State $\synccount \gets 0$
		\While {$\len(\signal)>\len(\chunks[0])+max\{\nextlen\}$}
			\State $j \gets \argmax_{j\in \nextlen}{\{\corr(\chunks[0], \signal[j:\clen])\}}$
			\State $c \gets \corr(\chunks[0], \signal[j:\clen])$
			\If {$\corr(\chunks[0], \signal[:j])>T$ \textbf{and} $\state=\STnormal$}
				\State $\chunks.\append(\signal[:j])$ // master chunk
			\Else
				\If {c < T}
					\State $\state \gets \STsync$
					\State $\nextlen \gets \left[S, S+6000\right]$
					\State $\synccount+=1$
					\If {$\synccount > 3$}
						\State return \STerror
					\EndIf
				\Else
					\State $\state \gets \STnormal$
					\State $\nextlen \gets G$
				\EndIf
			\EndIf
			\State $\signal \gets \signal[j:]$
		\EndWhile
		\BState \algstep{outlier rejection}:
		\State $\outlierreject(\chunks)$

		\State \textbf{return} $\chunks$
		\EndProcedure
	\end{algorithmic}
%COMMENT
\end{center}

%COMMENT

\subsubsection{Comparing against a natural baseline}
\label{sec:algocompare}

Figure~\ref{algocompare} compares our method with the baseline, correlation-based one outlined in Section~\ref{sec:chunkchall}, for processing a particularly noisy Zebra signal recorded using a parabolic mic directed at a Soyo screen. This naive approach comprises of (1) chunking according to the the exact de-facto refresh rate (59.9019\ in this case), (2) rotationally shifting chunks according to maximal correlation, and (3) performing outlier rejection, removing chunks whose correlation is less than 0.05.

\begin{figure*}[t]
\centering
	\vspace{-2.5em}
\subfloat[Output of the naive approach.]{
  \includegraphics[width=0.31\textwidth]{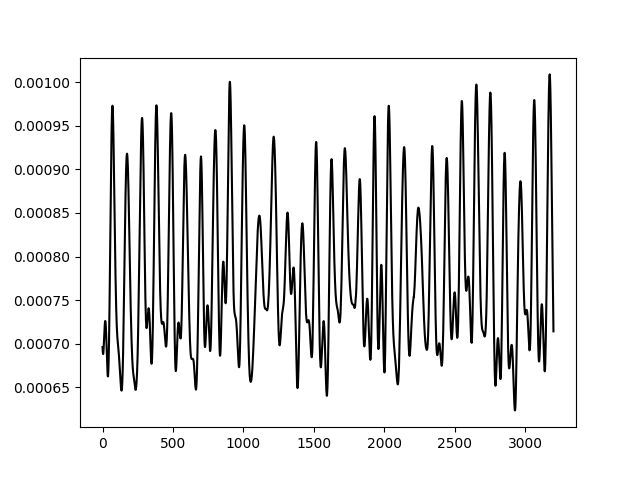}
  \label{naturalApproach}
}
\hfill
\subfloat[Output of our approach, using $S=2306,T=0.05,d=3$.]{
  \includegraphics[width=0.31\textwidth]{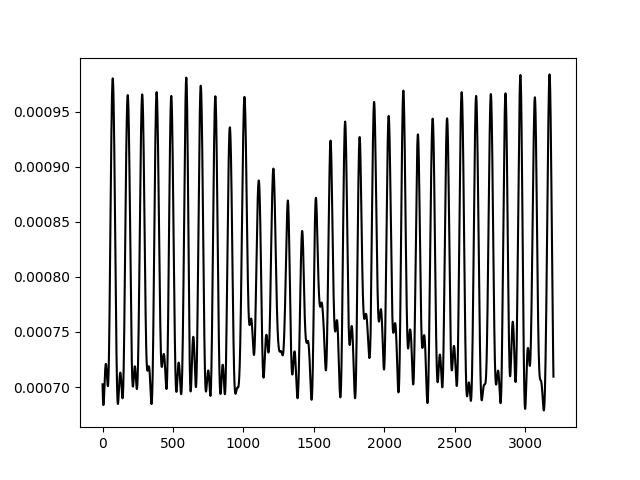}
  \label{notriggerOur}
}
\hfill
\subfloat[Output of ``ground truth'' chunking, using the trigger signal obtained via the vsync probe to accurately segment the trace.]{
  \includegraphics[width=0.31\textwidth]{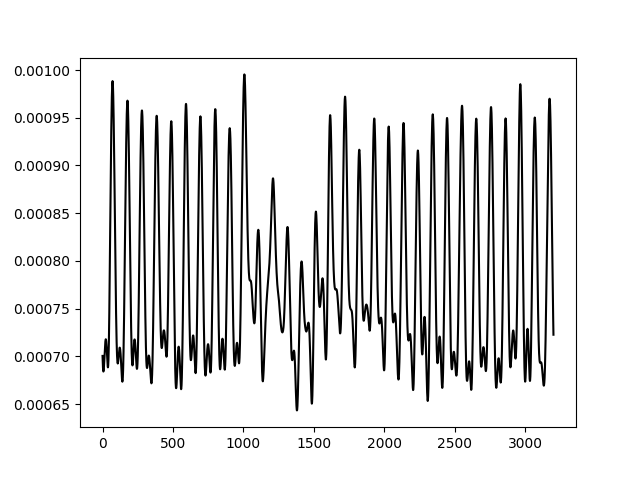}
  \label{triggerOur}
}
%COMMENT
	\caption{Comparing the outputs of the two chunking approaches on a Zebra of 30 periods, recorded using a parabolic mic (see Section~\ref{sec:closevsfar}). The naive approach produces a signal whose amplitude is erratic and bears less resemblance to the one of the ground truth, even though it is using as input the exact refresh rate (up to 0.0001\second), and even though this rate is entirely stable across this particular recording (there are no abnormally sized cycles).
\label{algocompare}}
\iftwocol{}
\vspace{-1.5em}
\fi{}
\end{figure*}

\ignore[Finding $S$]{
\subsubsection{Finding $S$ without the vsync probe}
\label{appx:bruteforce}
To find the correct value for $S$ (if it is unknown for the victim screen), we sample several victim recordings, and set $S=S'$ for every $S'\in \{3200, ..., 3209\}$\footnote{We use a small safety margin around the range we empirically measured across all of our settings, $\{3202,...,3206\}$.}, we run the chunking algorithm with $d=1$ for all recordings and measure its average \textit{score} across recordings. The score comprises the average chunk correlation, with a ``penalty'' for sync mode iterations (where correlation is not above the threshold). Concretely, the score is $\frac{s-m}{I}$, where $s$ is the sum of fetched chunk correlations, $m$ is the correlation threshold $T$ multiplied by the number of sync mode iterations, and $I$ is the total number of iterations. Recall the algorithm is designed to handle a drift of $d=1$ sample, so we expect the scores at 1-sample vicinity to the ground-truth $S$ to be the highest. Thus, we set $S$ as the value $S'$ that maximizes $average\{score(S'-1), score(S'), score(S'+1)\}$.

\parhead{Evaluation}
We used our recordings for the Website distinguishing attack from Section~\ref{modulatedwebsite}. These contain traces captured in three different settings: close range, phone, and at-distance.
%COMMENT
For each setting, we sampled 10 arbitrarily-chosen websites and one recording from each website. This totals 50 seconds of recording from each setting. In each setting, the brute-force approach returned 3206, which is exactly the closest value to the actual refresh rate (in sample numbers) as measured by the vsync probe in these settings.

Below is an example for the scores in the phone setting.
\lstset{%COMMENT
  basicstyle=\ttfamily,
  columns=fullflexible,
  emphstyle=\underline,
}
\scriptsize{}
\begin{lstlisting}[escapechar=@]
3200:0.1274845721533317
3201:0.1880864389194053
3202:0.4330874599527286
3203:0.5783651017505056
3204:0.6385812179470909
@\textbf{3205:0.7177944248570538}@
@\textbf{\underline{3206:0.8026286313468061}}@
@\textbf{3207:0.8027669481325654}@
3208:0.6743524112497232
3209:0.6076583373047286
\end{lstlisting}
\label{listing}
\normalsize{}

Note that the values 3205, 3206, and 3207, i.e., the values at 1-sample vicinity from the ground-truth, 3206, clearly attain the highest scores. The value 3207 is very close to that of 3206. Guessing 3207 is ``almost correct'': the actual refresh cycle length (in samples) is in [3206,3207] (but closer to 3206).
}

\subsection{Neural Network Architectures}
\subsubsection{On-screen keyboard snooping}
\label{sec:cnnsnoop}
For the experiment in Section~\ref{sec:keyboardsnoop1}, we used a convolutional layer with window size 12, 16 filters, and stride 1, followed by another convolutional layer with the same parameters but 32 filters, followed by a max-pooling layer with pool size 4, and one fully-connected layer with output size 512. The convolutional and FC layers have ReLU activations. The last layer is a softmax. We used an Adadelta optimizer with a categorical crossentropy loss, and a batch size of 64. We trained each network for 200 epochs. Each epoch takes around 3\second. The model was evaluated on the validation set after every epoch; the best-performing model was saved for testing.

\subsubsection{Text extraction}
\label{sec:cnntext}
For the experiment in Section~\ref{textattack}, we used 1 convolutional layer with window size 12, 64 filters, and stride 1, one max-pooling layer with size 2, and 1 fully-connected layer with output size 2048. The convolutional and FC layers have ReLU activations. The last layer is softmax. We used an SGD optimizer with learning rate of 0.01, norm-clipped at 1. We used a batch size of 16, and trained for 1000 epochs. After every epoch the model was evaluated on the validation set; the best-performing model was saved for testing.

\subsubsection{Website distinguishing}
\label{sec:cnnwebsite}
For the experiment in Section~\ref{modulatedwebsite}, we used 6 convolutional layers, with a max-pooling layer after every 2. All convolutions are of window size 24 and stride 1. For the first and second layer, we have 16 and 32 filters respectively, and 64 for the other four convolutional layers. The 10th layer is a fully connected layer with 512 outputs, followed by a 0.9 dropout layer. The first 10 layers have ReLU activations. The last layer is an FC layer with softmax activations. We used an SGD optimizer with a 0.01 learning rate, 0.1 gradient clipping, a categorical crossentropy loss, and batch size of 64. We trained each network for 800 epochs (about 4\second per epoch).

For the experiment in Section~\ref{codecattack}, we used 3 convolutional layers with kernel size 12, stride 1 and 16 filters, followed by a max-pooling layer with pool size 8, followed by a fully connected (\textit{FC}) layer with output 512, followed by a Dropout 0.5 layer, followed by a softmax output layer. Convolutional, FC, and max-pooling layers have ReLU activations. We used an Adadelta optimizer with categorical crossentropy and batch size 16. We trained for 100 epochs. Each epoch took about 7\second.

\subsection{List of Screens Used} \label{sec:screenlist}
We specify the screen models used in the experiments reported in this paper. Similar effects were observed during opportunistic ad-hoc tests on additional screens.

\iftwocol{}
{\footnotesize
\else{}
\vspace{1em}
{
\fi{}
%COMMENT
\centering
\begin{tabular}{|l|c|c|c|c|}
\hline
 \textbf{Model} & \textbf{Size} & \textbf{Resolution} & \textbf{Backlight}  & \textbf{Qty} \\
\hline
Apple Cinema A1316 & 27" & 2560 $\times$ 1440  & LED & 1 \\
\hline
BenQ q20ws & 20.1" & 1680  $\times$ 1050  & CCFL & 1  \\
\hline
ViewSonic  VA903b & 19" & 1280 $\times$ 1024 & CCFL & 2  \\
\hline
Samsung   920NW & 19" & 1440 $\times$ 900  & CCFL & 1 \\
\hline
Dell 2208WFPt  & 22" & 1680  $\times$ 1050  & CCFL & 8  \\
\hline
Dell 2208WFPf  & 22" & 1680  $\times$ 1050  & CCFL & 2  \\
\hline
Dell 3011 & 30" & 2560 $\times$ 1600  & CCFL & 5 \\
\hline
Dell U2713H  & 27" & 2560  $\times$ 1440  & LED & 2  \\
\hline
HP ZR30w & 30" & 2560 $\times$ 1600  & CCFL & 5  \\
\hline
Philips 170S4 & 17" & 1280 $\times$ 1024  & CCFL & 1  \\
\hline
Soyo DYLM2086 & 20" & 1440 $\times$ 900 & CCFL & 1 \\
\hline
Soyo DYLM2248  & 22" & 1680  $\times$ 1050  & CCFL & 2 \\
\hline
Eyoyo S801C  & 8" & 1024 $\times$ 768 & LED & 1 \\
\hline
Eyoyo 808H  & 8" & 1024 $\times$ 768 & LED & 1 \\
\hline
\begin{tabular}{@{}l@{}}Lenovo Carbon X1\\3rd Gen laptop screen\end{tabular}
& 14" & 2650 $\times$ 1440  & LED & 1 \\
\hline

\end{tabular}
}

\subsection{The effect of microphone distance}
\label{sec:distancevsquality}
We investigate the effect of microphone proximity to the screen on the leakage signal. We used the setup in Section~\ref{background} to record a Dell 2208WFPt screen from various distances. The microphone was placed near the screen's top and then moved away along a straight line, pitch 40 degrees up, while maintaining the microphone's orientation towards the screen coaxially. We recorded traces of a Punctured Zebra from a distance of 1cm, 2cm, 5cm, 10cm, 20cm, 50cm, 100cm, 200cm, 300cm, and 500cm. Then, to measure the signal quality at a given distance, we performed the correlation test from Section~\ref{signalprocessing} and Figure~\ref{fig:triggered-blackhole}: we demodulated and ``chopped'' the signal into chunks (each corresponding with a refresh period) using the vsync probe. We then measured the average Pearson correlation of the chunks from the average of chunks.

%COMMENT
Figure~\ref{fig:distancevsqualityqual} shows the resulting chopped-and-averaged trace for each distance. Especially for small distances, one can readily observe the regular patterns corresponding to Zebra stipes, as well as the flatter region corresponding to the puncturing (black rectangle) in the middle. Figure~\ref{fig:distancevsqualityquant} shows the correlation values for the various distances. Even at a distance of 3 meters, we see nontrivial correlation value.

%COMMENT
\eran{SANITIZED}

Note the discernible shifts between the traces in Figure~\ref{fig:distancevsqualityqual}, as distance increases. These occur since the signal is captured acoustically but triggered electronically. At a 60\kHz refresh rate, the speed of sound (343\meter/s) causes a delay of 17.5\% of the refresh cycle per meter of distance --- consistently with the observed signal. Had the signal source been electromagnetic (conducted or emanated), the delay would have been induced by the speed of light in the relevant medium (metal or air), and thus smaller by 6 orders of magnitude.

\begin{figure}[t]
\vspace{-2em}
\centering
	\subfloat[For various distances, the averages of modulated signal segments corresponding with refresh periods.]{
		\includegraphics[width=0.85\columnwidth]{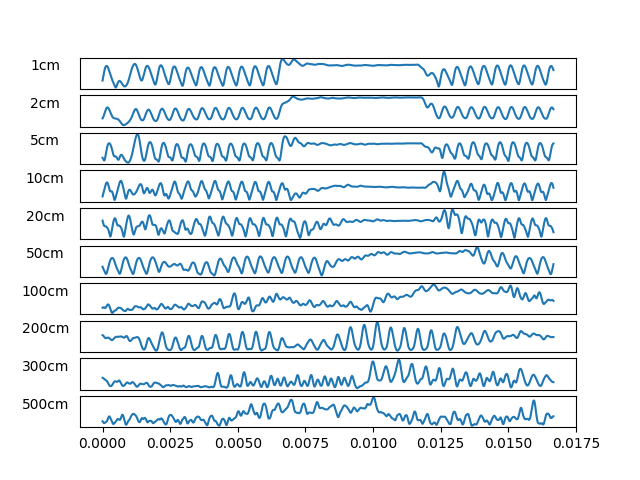}
		\label{fig:distancevsqualityqual}
	}
\vspace{-1em}
	\subfloat[Signal quality (average correlation value) as a function of microphone distance.]{
		\includegraphics[width=0.85\columnwidth]{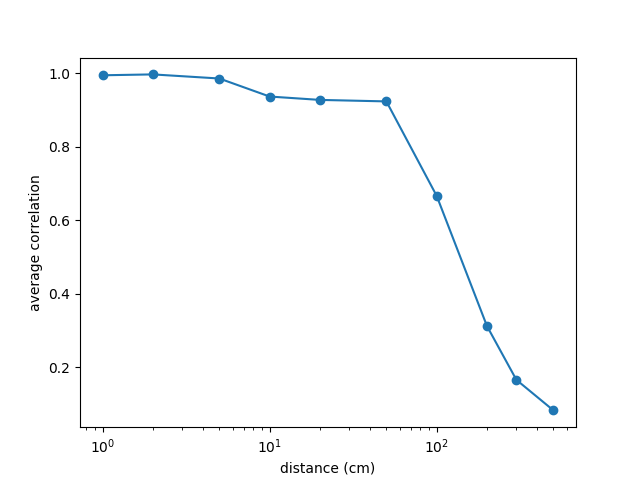}
		\label{fig:distancevsqualityquant}
	}
\caption{
\label{fig:distancevsquality}
	For recordings of a black hole from various distances, we quantitatively and qualitatively characterize the relationship between microphone distance and signal quality.
}
\vspace{-1em}
\end{figure}

\done\eran{SANITIZED}

%COMMENT
\end{document}